\documentclass[usenatbib,usegraphicx,letterpaper]{mn2e}
\usepackage[totalwidth=480pt,totalheight=680pt]{geometry}

\usepackage{amssymb}
\usepackage{epsfig}
\usepackage{amsmath}
\usepackage{color}
\usepackage[dvipsnames]{xcolor}
\usepackage{yfonts}

\usepackage{epsfig}  
\usepackage{graphicx}  
\usepackage{rotating}

\newcommand{\lsim}{\lower0.6ex\vbox{\hbox{$ \buildrel{\textstyle <}\over{\sim}\ $}}}
\newcommand{\gsim}{\lower0.6ex\vbox{\hbox{$ \buildrel{\textstyle >}\over{\sim}\ $}}}
\newcommand{\beq}{\begin{equation}}
\newcommand{\eeq}{\end{equation}}

\newcommand{\mvir}{M_{\rm vir}}

\newcommand{\rvir}{R_{\rm vir}}
\newcommand{\ncen}{N_{\rm cen}}
\newcommand{\nsat}{N_{\rm sat}}
\newcommand{\ngal}{N_{\rm gal}}
\newcommand{\abias}{\mathcal{A}_{\rm bias}}

\def\gtsima{$\; \buildrel > \over \sim \;$}
\def\ltsima{$\; \buildrel < \over \sim \;$}
\def\prosima{$\; \buildrel \propto \over \sim \;$}
\def\gsim{\lower.7ex\hbox{\gtsima}}
\def\lsim{\lower.7ex\hbox{\ltsima}}
\def\simgt{\lower.7ex\hbox{\gtsima}}
\def\simlt{\lower.7ex\hbox{\ltsima}}
\def\simpr{\lower.7ex\hbox{\prosima}}

\def\ga{\gsim}

\def\gta{\ga}

\newcommand{\wprp}{w_{\mathrm{p}}}
\newcommand{\rp}{r_{\mathrm{p}}}
\newcommand{\magr}{M_r}

\newcommand{\bit}{\begin{itemize}}
\newcommand{\eit}{\end{itemize}}
\newcommand{\ben}{\begin{enumerate}}
\newcommand{\een}{\end{enumerate}}

\title[Clustering Constraints on Assembly Bias]
{
Constraints on Assembly Bias from Galaxy Clustering
}

\author[Zentner et al.]
{Andrew R. Zentner$^{1}$, Andrew Hearin$^{2}$, Frank C. van den Bosch$^{3},$ \newauthor
Johannes U. Lange$^{3}$, and Antonio Villarreal$^{1}$\\ \\
$^1$Department of Physics and Astronomy \& Pittsburgh Particle Physics, Astrophysics, and Cosmology Center (PITT PACC),\\ University of Pittsburgh, Pittsburgh, PA 15260\\
$^2$Yale Center for Astronomy \& Astrophysics, Yale University, New Haven, CT\\
$^3$Department of Astronomy, Yale University, P.O. Box 208101, New Haven, CT\\
}

\date{\today}

\pagerange{\pageref{firstpage}--\pageref{lastpage}} \pubyear{}

\begin{document}

\maketitle
%----------------------------------------------------------------
%%%%%%%%%%%%%%%%%%%%%%%  A B S T R A C T %%%%%%%%%%%%%%%%%%%%%%%%%%%%%%
\begin{abstract}
  We constrain the newly-introduced decorated Halo Occupation Distribution (HOD) 
  model using SDSS DR7 measurements of projected galaxy clustering, $\wprp (r_{\rm p})$ 
  of galaxies in $r$-band luminosity-threshold samples. 
  The decorated HOD is a model for the galaxy--halo
  connection that augments the traditional HOD by allowing for the possibility of 
  {\em galaxy assembly bias}: galaxy luminosity may be correlated with dark matter halo
  properties besides mass, $\mvir$. We demonstrate that it
  is not possible to rule out galaxy assembly bias using DR7 measurements of galaxy 
  clustering alone. Moreover, galaxy samples $\magr<-20, -20.5$
  favor strong levels of central galaxy assembly bias. These samples prefer scenarios  
  in which high-concentration halos are more likely to host a central galaxy
  relative to low-concentration halos of the same $\mvir.$ We rule out
  zero assembly bias with high significance for these
  samples. Satellite galaxy assembly bias is significant for the
  faintest sample we study, $\magr<-19.$ We find no evidence for
  assembly bias in the $\magr<-21$ sample. Assembly bias should be accounted for 
  in galaxy clustering analyses or attempts to exploit galaxy clustering to constrain cosmology. 
  In addition to presenting the first constraints on HOD models that accommodate assembly bias, our analysis 
  includes numerous improvements over previous analyses of this data set and 
  supersedes previously published results, even in the case of a standard HOD analysis.
\end{abstract}

%---------------------------
\section{Introduction}
\label{section:introduction}
%---------------------------

For more than a decade, halo occupation modeling has been used to
interpret large-scale structure measurements and exploit these
measurements to constrain galaxy formation models and cosmology
\citep[e.g.,][]{yang03,tinker05,zehavi05a,
  porciani06,vdBosch07,Zheng07,conroy_wechsler09,yang09b,zehavi_etal11,guo_etal11b,
  wake_etal11,yang11a,yang12,leauthaud_etal12,rod_puebla12,tinker_etal13,cacciato_etal13,
  more_etal13,guo_etal14,zu_mandelbaum15b}. The key assumptions
underlying halo occupation modeling are: (1) all galaxies reside in
dark matter halos that are biased tracers of the density field; and
(2) galaxies occupy halos as a function of halo mass $\mvir$ only. 
In this paper, we present the first analysis of galaxy clustering using 
a halo occupation model that violates the second of these assumptions 
and permits galaxies to occupy halos as a function of multiple halo properties.

It is now well known that the strength of halo clustering, the halo bias, 
depends upon halo properties other than $\mvir$
\citep[e.g.][]{gao_etal05,wechsler06,gao_white07,zentner07,dalal_etal08,lacerna11},
an effect called {\em halo assembly bias}. If galaxies occupy halos as a
function of halo properties other than $\mvir,$ then standard halo
occupation methods will be subject to a systematic error due to 
{\em galaxy assembly bias}. Several of us have previously shown 
that this error can be significant in an analysis of galaxy clustering 
and can bias inferences about many aspects of galaxy evolution \citep{zentner_etal14}.

There is increasing observational evidence that galaxy assembly bias is
present in the real Universe.
Using an approach based on subhalo abundance matching,
\citet{lehmann_etal15} showed that the clustering of low-redshift
galaxies favors models in which stellar mass depends upon a
combination of $\mvir$ and halo concentration.
\citet{miyatake_etal16} and \citet{more_etal16} have
presented evidence for the presence of assembly bias in massive
clusters; in particular, they find that clusters with a more centrally concentrated 
distribution of satellite galaxies cluster more weakly than clusters with more 
diffuse satellite galaxy distributions at fixed cluster mass. Indeed, \citet{more_etal16} 
suggest that the observed galaxy bias is considerably stronger than had been 
expected based on $N$-body simulations coupled with simple models for the 
galaxy--halo connection, such as subhalo abundance matching. 
Additional support for the presence of assembly bias has come from a
variety of studies that have shown, or suggested, that the large-scale
environment of dark matter halos of fixed mass
is correlated with the star formation rate of their central galaxies
\citep{yang_etal06a, blanton_berlind07,wang_etal08,wang_etal13,hearin_etal14}.

Despite this growing evidence, the empirical modeling community
has not yet reached a consensus on the question of galaxy assembly bias.
For example, \citet{lin_mandelbaum_etal15} claim that
some of the above evidence for assembly bias can be explained by
differences in halo mass between the galaxy samples under consideration.
\citet{paranjape15} urges caution that contamination by satellite galaxies
can masquerade as a false signal of assembly bias.
\citet{tinker08b} argue that measurements of void statistics place strong
bounds on the possible strength of the signal, though 
\citet{zentner_etal14} constructed an explicit example refuting this 
claim.

In much of the literature on this topic,
the character of the supporting evidence
suffers from a severe shortcoming.
Until very recently, the only models that have been fit to
observational clustering measurements have been traditional models
in which assembly bias is assumed from the outset to be zero.
Models that include assembly bias such as semi-analytic models
are typically too computationally expensive to compare to data in a manner 
that enables statistical constraints on model parameters 
\citep[but see][for recent progress in this direction]{lu_etal11,lu_etal12,henriques_etal13,lu_etal14}. 
Such models are usually 
compared to data on a case-by-case basis \citep[e.g.,][]{croton_etal07}, prohibiting
any conclusive statement to be made about the strength of any assembly bias signal.

Motivated by this shortcoming, we have developed a new class of
empirical models that enable galaxies to occupy halos in a manner
that has {\em simultaneous} dependence
upon multiple halo properties \citep{hearin_etal16},
{\em including continuously variable levels of assembly bias}.
Crucially, our implementation is of sufficient computational efficiency
to permit a proper likelihood analysis of the model parameter space.
Armed with this new methodology, in this paper
we revisit the interpretation of luminosity-dependent
galaxy clustering in the Sloan Digital Sky Survey (SDSS) Data Release
7 (DR7) data, analyzed previously by \citet{zehavi_etal11}, in the
context of both standard halo occupation models and these new models 
that permit parameterized galaxy assembly bias.

This work presents a re-analysis of the SDSS DR7 data that
overcomes numerous shortcomings of previous studies.
The primary improvements of our approach stem from the fact that we 
populate directly halos within a cosmological simulation with mock galaxies, 
whereas previous analyses relied upon analytical fitting functions
with parameters calibrated against a suite of simulations. 
In direct mock population, delicate issues present in analytic
modeling, such as scale-dependent halo bias and halo exclusion, 
are treated exactly. 
Our approach also provides a natural framework for studying 
models of assembly bias, because we can use the exact clustering 
properties of simulated halos as a function of
any arbitrary halo properties in the catalog. It is not necessary to develop 
new phenomenological models or fitting functions to deal with this additional 
complexity. With direct mock population, systematic uncertainty in the 
model is limited to the sample variance of the simulated box,
numerical inaccuracies of halo-finding, and errors related 
to insufficient numerical resolution \citep[e.g., overmerging; ][]{klypin99a,rockstar}.

In addition to improvements resulting from direct mock population,
the simulation we use is based on the latest Planck cosmological
parameters \citep{planck13}, updating previous work.
Furthermore, many important differences between this work 
and the previous analysis by \citet{zehavi_etal11} result from 
the fact that the Monte Carlo Markov Chains used in \citet{zehavi_etal11} did not 
sufficiently sample the posterior distribution (Z. Zheng, private communication). 
The result is that the constraints in \citet{zehavi_etal11} are overly restrictive, considerably so 
in some cases, and potentially biased. 
Therefore, {\em our analysis supersedes previous work even in the case of
standard halo occupation models}.

Most importantly, our work demonstrates explicitly that significant
assembly bias in $\magr$-selected samples from SDSS DR7 cannot be ruled
out based on a standard analysis of galaxy clustering only. In fact,
in agreement with \citet{lehmann_etal15}, we find that several samples
{\em favor galaxy assembly bias to a degree that is statistically significant}. 
As we were completing this work, we became aware of an independent study 
using similar techniques and reaching broadly similar conclusions \citep{vakili_etal16}. 
As demonstrated by \citet{zentner_etal14}, the assembly bias suggested by the 
clustering of these samples has important consequences for the interpretation of
both extant and forthcoming data as well as predictions for independent statistics 
that make use of the HODs derived via clustering.

Our paper is organized as follows. In Section~\ref{section:methods},
we discuss our implementation of halo occupation models and the
parameter inference methodology. We present results
from both traditional and assembly-biased halo occupation analyses
in Section~\ref{section:results}. We discuss and interpret our results 
in Section~\ref{section:discussion}. We summarize our results and draw
conclusions in Section~\ref{section:conclusions}.

%---------------------------
\section{Methods}
\label{section:methods}
%---------------------------

%---------------------------
\subsection{Halotools Implementation of HOD Models}
\label{subsection:halotools}
%---------------------------

To generate predictions for galaxy clustering, 
we populate dark matter halos with mock galaxies using the publicly-available, 
open source, {\tt Halotools} software \citep{halotools}.
We explore halo occupation distribution (HOD) models in this work
\citep[e.g.][]{seljak00,ma_fry00,scoccimarro01a,berlind02}, though other
techniques that can be used to interpret such data, such as the conditional luminosity function
\citep[CLF, e.g.,][]{yang03,vdBosch13}, exist. In this subsection, we review the ``standard" HOD
model used in the present work, which assumes that there is no galaxy assembly bias. We refer to
such a model as ``standard" because all HOD analyses of galaxy clustering to date have assumed no
galaxy assembly bias. In the following subsection, we describe
the Decorated HOD model described in \citet{hearin_etal16}. In both cases, we will only
review the salient features of our methodology briefly;
interested readers can always refer to {\tt http://halotools.readthedocs.io} and
\citet{hearin_etal16} for further details.

\subsubsection{Simulation}

All of our analyses are based on the Bolshoi-Planck (BolshoiP) simulation \citep{riebe_etal11}.
BolshoiP was run with cosmological parameters based on \citet{planck13}:
$\Omega_{\Lambda} = 0.693; \Omega_{\rm m} = 1 - \Omega_{\Lambda} = 0.307; \Omega_{\rm b} = 0.048;
h = 0.7; {\rm n_s}=0.96$; and $\sigma_8 = 0.82$. BolshoiP simulated the formation of structure within a cubic 
box $250$ ${\rm Mpc}/h$ on a side, requiring a particle mass of $m_{\rm p} = 1.35\times10^{8}M_{\odot}/h.$
Further information about the BolshoiP simulation is available at {\tt https://www.cosmosim.org}.

We use publicly available\footnote{\tt http://www.slac.stanford.edu/$\sim$behroozi/BPlanck\_Hlists} dark matter
halo catalogs based on the {\tt ROCKSTAR} halo-finder \citep{behroozi_rockstar11} 
and {\tt CONSISTENT TREES} algorithm \citep{behroozi_trees13}. In particular, we use the {\tt halotools\_alpha\_version2} 
version of the $z=0$ snapshot
of the {\tt `bolplanck'} catalog included with {\tt Halotools.} Halos in these catalogs are defined by the virial 
radius density contrast given in \citet{bryan_norman98} and have virial masses $\mvir$ within their virial 
radii. When populating this catalog with mock galaxies, we only use
present-day host halos with a value of $M_{\rm peak}$ that exceeds $300$ particles, where $M_{\rm peak}$ is 
the maximum mass a halo obtains during its evolution. We consider only host halos and not their 
subhalos in this work. For host halos, $M_{\rm peak}$ is almost always nearly identical to the 
present day virial mass of the halo, $\mvir$. The minimum 
peak halo mass considered in our analysis has $M_{\rm peak} \ge 4.05 \times 10^{10}\, h^{-1}\mathrm{M}_{\odot}$ 
and this prevents us from analyzing samples with $M_r > -19$. Throughout the remainder of this 
paper, we will refer only to the virial mass $\mvir$ of the halo to be in closer concordance with other 
work on HOD analyses.

\subsubsection{Occupation statistics}

In standard HOD models, central galaxies and satellite galaxies are treated separately, so
the model is specified by two probability distributions, one for each type of galaxy.
The galaxy-halo connection is specified in terms of $P(\ncen|\mvir)$ and $P(\nsat|\mvir),$
the probability that a halo of mass $\mvir$ hosts $\ncen$ central and $\nsat$ satellite galaxies,
respectively. $P(\ncen|\mvir)$ is typically a nearest-integer distribution, as a host halo has only
either zero or one central galaxy. Consequently, the occupation statistics of central galaxies are
specified by the first moment of $P(\ncen|\mvir)$, which we model as
\begin{equation}
\label{eq:ncen}
\langle N_{\mathrm{cen}} \vert \mvir \rangle =
        \frac{1}{2}\left( 1 +
        \mathrm{erf}\left[ \frac{\log (\mvir) -
        \log (M_{\rm min})}{\sigma_{\log M}} \right] \right)
\end{equation}
For every host halo in the catalog we draw a random number from a uniform 
distribution $\mathcal{U}(0, 1);$ for a host halo of present-day virial mass $\mvir,$ 
a central galaxy is assigned to the halo if the associated random number is less 
than $\langle N_{\mathrm{cen}} \vert \mvir \rangle;$ halos with random values 
exceeding  $\langle N_{\mathrm{cen}} \vert \mvir \rangle$ are left devoid of centrals.
The parameter $\log (M_{\rm min})$ specifies the halo mass at which the halo 
has a 50\% probability of hosting a
central galaxy, while the parameter $\sigma_{\log M}$ specifies the rate at 
which $\langle N_{\mathrm{cen}} \vert \mvir \rangle$
transitions from zero to unity, with smaller values of $\sigma_{\log M}$ 
corresponding to a more rapid transition.

We model the distribution $P(\nsat|\mvir)$ as a Poisson distribution with first moment given by a power law,
\begin{equation}
\label{eq:nsat}
\langle N_{\rm sat}\vert \mvir \rangle = \left( \frac{\mvir - M_0}{M_1} \right)^{\alpha}.
\end{equation}
The parameter $M_0$ allows the power-law to be truncated more rapidly at low masses and we
set $\langle N_{\rm sat}\vert \mvir \rangle = 0$ for $\mvir < M_0$.

The five parameters of HOD models that are varied in standard analyses are $\log (M_{\rm min})$,
$\sigma_{\log M}$, $\alpha$, $\log (M_1)$, and $\log (M_0)$, though, as we show below, central galaxies usually
outnumber satellite galaxies by a factor of several, so $\log (M_{\rm min})$ and $\sigma_{\log M}$ usually
vary along a narrow degeneracy that fixes the total galaxy number density to the observed value. There are
many particular choices that can be made for the functional forms of $\langle N_{\rm cen}\vert \mvir \rangle$ and
$\langle N_{\rm sat}\vert \mvir \rangle$. We have made choices that mimic the standard SDSS DR7 analysis of
\citet{zehavi_etal11}, to expedite comparisons with their results. However, our choice does deviate from 
\citet{zehavi_etal11} in one respect. The mean satellite occupation of \citet{zehavi_etal11} 
is that of Eq.~(\ref{eq:nsat}) multiplied by an overall factor of $\langle N_{\rm cen} \vert \mvir \rangle$. 
We have chosen not to use this as our default because it can introduce difficulties in some analyses 
\citep[see the discussion of blue galaxy samples in][]{zentner_etal14} and because, as we have verified 
explicitly, the extra factor introduces only small quantitative changes to our results and no qualitative 
changes.

\subsubsection{Galaxy profiles}

Central galaxies in the standard HOD models reside at the halo center,
moving with the same velocity as the host halo peculiar velocity. We
model the intra-halo spatial distribution of satellite galaxies to be
located within $\rvir$ of the halo center, with a spherically
symmetric NFW profile \citep{nfw97}.  The concentration $c$ of each
halo's satellite galaxy profile is taken to be the same as the
concentration of the dark matter particles in the halo.\footnote{We
  set a maximum value of $c=25$ to the NFW concentration, because
  halos with very large values for the concentration tend to be poorly
  described by an NFW profile, for example due to a recent merger.}

We model the radial velocity distribution of satellite galaxies as a
Gaussian with first moment equal to the host halo velocity and second
moment equal to the solution of the isotropic Jeans equation for an
NFW profile \citep{more09b},
\begin{equation}
\sigma^{2}_{r}(\tilde{r}|c) = V_{\rm vir}^{2}\frac{c^{2}\tilde{r}(1 + c\tilde{r})^{2}}{g(c)}\int_{c\tilde{r}}^{\infty}{\rm d}y\frac{g(y)}{y^{3}(1 + y)^{2}},
\end{equation}
where $\tilde{r} = r/\rvir,$ $g(x) = \rm{ln}(1+x) - x / (1+x),$ and
$V_{\rm vir}^{2} = G\mvir/\rvir.$ We assume that velocities are
isotropic, setting the peculiar velocities in each Cartesian direction
according to random draws from the above radial velocity distribution.

\subsubsection{Predictions for observables}

After populating a halo catalog with mock galaxies, we calculate the
comoving number density of our mock galaxy sample as $n_{\rm g} =
N_{\rm gal} / L_{\rm box}^3$, where $N_{\rm gal}$ is the total number of
galaxies in the mock sample. We apply the distant-observer
approximation and use the simulation $z$-axis as the line-of-sight
direction, and the distance between points in the $xy$-plane to define
the projected distance $\rp.$ We place mock galaxies into
redshift-space by replacing each galaxy's $z$-coordinate with
$z_{\rm RS} = z + V_{\rm z}/H_0$\footnote{A similar exercise is demonstrated 
as part of the Halotools documentation at {\tt http://halotools.readthedocs.io}}. 
Having populated mocks, 
we perform the computation of the projected two-point galaxy correlation 
function, $w_{\rm p}(r_{\rm p})$ using the publicly-available {\tt CorrFunc} 
package \citep{corrfunc} which has been extensively optimized for computational speed.
We count pairs of points in each of
our $\rp$ bins, rejecting pairs with line-of-sight distances in redshift space, $\Delta z_{\rm RS}$, 
exceeding $\pi_{\rm max}=60$ ${\rm Mpc/h}$, which is the same projection
depth chosen by \citet{zehavi_etal11}, both in their analysis of the SDSS
DR7 data and in their modeling. 

%---------------------------
\subsection{HOD with Assembly Bias: The Decorated HOD}
\label{subsection:decorated}
%---------------------------

In addition to the standard occupation statistics described in the
previous section, in this paper we also use the decorated HOD
formalism to connect galaxies to dark matter halos in a manner that
has simultaneous dependence on both $\mvir$ and {\em halo concentration}.
Briefly, we use Equations \ref{eq:ncen} and \ref{eq:nsat} as our
``baseline" first occupation moments.  At fixed $\mvir,$ halos are
divided into one of two categories, those of high- and
low-concentration, depending on whether the concentration of the halo
places it above or below the rank-order percentile $f_{\rm split},$
which we keep fixed to $f_{\rm split}=0.5$ throughout the paper for
simplicity. High-concentration halos have a different first
occupation moment relative to low-concentration halos of the same
mass,
\begin{eqnarray}
\langle\ngal|\mvir, c_{\rm high}\rangle & = & \langle\ngal|\mvir \rangle + \delta\ngal \nonumber \\  
\langle\ngal|\mvir, c_{\rm low}\rangle  & = & \langle\ngal|\mvir \rangle - \delta\ngal
\end{eqnarray}
This difference, $\delta\ngal$, between the
first moment of high- and low-concentration halos is modulated by
$\abias,$ the novel parameter of the decorated HOD governing assembly
bias. Values of $\abias=\pm1$ correspond to the maximum strength of
assembly bias allowable by the constraint that the model preserves the
marginalized first moment, $\langle\ngal|\mvir\rangle;$ thus
regardless of the value of $\abias,$ in the decorated HOD the
marginalized first moment of centrals and satellites are {\em 
unchanged from the baseline value defined by Equations (\ref{eq:ncen}) 
and (\ref{eq:nsat})}. In other words, {\em decorated HOD models all have the same HODs,
when averaged over all halos at fixed mass, as standard HOD
models with the same five standard HOD parameters}. 
The only change in Decorated HOD models is whether or not an
additional property also modulates halo occupation at fixed halo
mass. A value of $\abias = 0$ indicates no galaxy assembly bias
whatsoever. We refer the reader to \citet{hearin_etal16} for further
details about the decorated HOD.

In the present work, we fix our model to the simplest class of galaxy
assembly bias models, though a recipe for generalizing to more
complicated models can be found in \citet{hearin_etal16}. In
particular, we split halos into two populations as specified in the
previous paragraph. We then populate halos with satellite galaxies
specified by an assembly bias parameter $-1 \le A_{\rm sat} \le 1$ and
central galaxies with a distinct assembly bias parameter
$-1 \le A_{\rm cen} \le 1$. For the value of $f_{\rm split}=0.5$ adopted here,
we then have that
\begin{eqnarray}
\delta N_{\rm cen} & = & A_{\rm cen} \, {\rm MIN} \left[ \langle N_{\mathrm{cen}} \vert \mvir \rangle ,
1- \langle N_{\mathrm{cen}} \vert \mvir \rangle \right]  \nonumber \\ 
\delta N_{\rm sat} & = & A_{\rm sat} \, \langle N_{\mathrm{sat}} \vert \mvir \rangle
\end{eqnarray}
In our analyses that include assembly bias, we vary these two
additional parameters, bringing the total number of parameters
that vary in these analyses to seven. As we show below, these additional
parameters are in many instances poorly constrained by clustering data
of the quality of SDSS DR7 alone, so exploring more complex models of
assembly bias does not yet seem justified in such analyses.

%---------------------------
\subsection{Parameter Inference}
\label{subsection:mcmc}
%---------------------------

We constrain HOD parameters based on SDSS DR7 measurements of the projected
galaxy two-point functions, $w_{\rm p}(r_{\rm p})$, and galaxy number densities, $n_{\rm g}$,
for luminosity threshold samples published in \citet{zehavi_etal11}. We use the full covariance
matrix of the projected correlation function available at
{\tt astroweb.cwru.edu/izehavi/dr7\_covar/table8}. We assume a likelihood of the
form $\mathcal{L} \propto e^{-\chi^2/2}$, where
\begin{equation}
\label{eq:chisquare}
\chi^2 = \Delta w_{{\rm p},i} \, [C^{-1}]_{ij} \, \Delta w_{{\rm p},j} + \frac{(n_{\rm g}^{\rm mock} - n_{\rm g}^{\rm meas})^2}{\sigma_{\rm n}^2},
\end{equation}
$\Delta w_{{\rm p},i} = w_{\rm p}^{\rm mock}(r_{{\rm p},i}) - w_{\rm p}^{\rm meas}(r_{{\rm p},i})$ is the
difference between the projected two-point function predicted by the
mock catalog and the measured value in the i$^{\rm th}$ separation bin
(of 11 bins total), $C^{-1}$ is the inverse of the covariance matrix
of the measurements, and repeated indices are summed over. 
The last term in Eq.~(\ref{eq:chisquare}) is the contribution from the difference 
between the predicted and measured galaxy number densities. 
We consider only the eleven values of 
$w_{\rm p}(r_{\rm p})$ given by \citet{zehavi_etal11} in bins logarithmically spaced between 
$r_{\rm p}=0.17\, h^{-1}\mathrm{Mpc}$ and $r_{\rm p}=16.9\, h^{-1}\mathrm{Mpc}$. Though 
\citet{zehavi_etal11} quote values at two larger separations, we find that considering those additional 
data points adds considerably to the computational expense but alter our results insignificantly. 
The error on the galaxy number density assumes Poisson statistics for both the measured and 
predicted galaxy number densities.

To infer parameters for the HOD and Decorated HOD models described in the previous subsections,
we perform a Markov Chain Monte Carlo (MCMC) sampling of the posterior distribution using the
affine-invariant ensemble sampler of \citet{goodman_weare10} as implemented in the
{\tt emcee} software package \citep{foreman-mackey_etal13}. For most cases, we find that
$\sim 3-10 \times 10^{6}$ samples are necessary in order for our chains to converge.

%-------------------------------------------------------------------------------------------------------------------------------------------------
\begin{table}
\begin{center}
{\renewcommand{\arraystretch}{1.3}
\renewcommand{\tabcolsep}{0.2cm}
\begin{tabular}{l c}
\hline
\hline
Parameter & Prior Interval\\
\hline
$\log (M_{\mathrm{min}})$ & [9.0,14.0] \\
$\sigma_{\log M}$ & [0.01,1.5] \\
$\log (M_0)$ & [9.0,14.0]\\
$\log (M_1)$ & [10.7,15.0]\\
$\alpha$ & [0.0,2.0]\\
$A_{\mathrm{cen}}$ & [-1.0,1.0]\\
$A_{\mathrm{sat}}$ & [-1.0,1.0]\\
\hline
\end{tabular}
\medskip
\caption{
Ranges for the priors used in the parameter inference. All prior distributions are uniform over the
specified ranges.}
 }
 \label{table:priors}
 \end{center}
\end{table}
%--------------------------------------------------------------------------------------------------------------------------------

The most important detail of this analysis is the priors on the parameters. In all analyses
discussed in this paper, we adopt priors that are uniform distributions over the intervals
specified in Table~1. In the case of the assembly bias parameters
$A_{\mathrm{cen}}$ and $A_{\mathrm{sat}}$, the priors have strict
boundaries. Mathematically, these parameters must satisfy $-1 \le A_{\mathrm{cen,sat}} \le 1$.
Physical considerations require parameter $\sigma_{\log M} > 0$. All other
priors have a negligible influence on the posterior aside from $\log M_0$ and 
$\sigma_{\log M}$. We find that $\log M_0$ is often very poorly constrained by clustering data and
priors on $\log M_0$ can have a non-negligible influence on inferred parameters. The parameter 
$\sigma_{\log M}$ is poorly constrained for the $\magr < -19$ sample.

%---------------------------
\section{Results}
\label{section:results}
%---------------------------

We have performed parameter inference analyses in order to infer the underlying HODs of
galaxies from the projected galaxy two-point function $\wprp(\rp)$ as described in the preceding
section. In this section, we describe the primary results of these analyses. Our marginalized
one-dimensional parameter constraints are given in Table~2.

%-------------------------
\subsection{Standard Analysis}
\label{subsection:standard}
%-------------------------

\begin{figure*}
\begin{center}
\includegraphics[width=15.0cm]{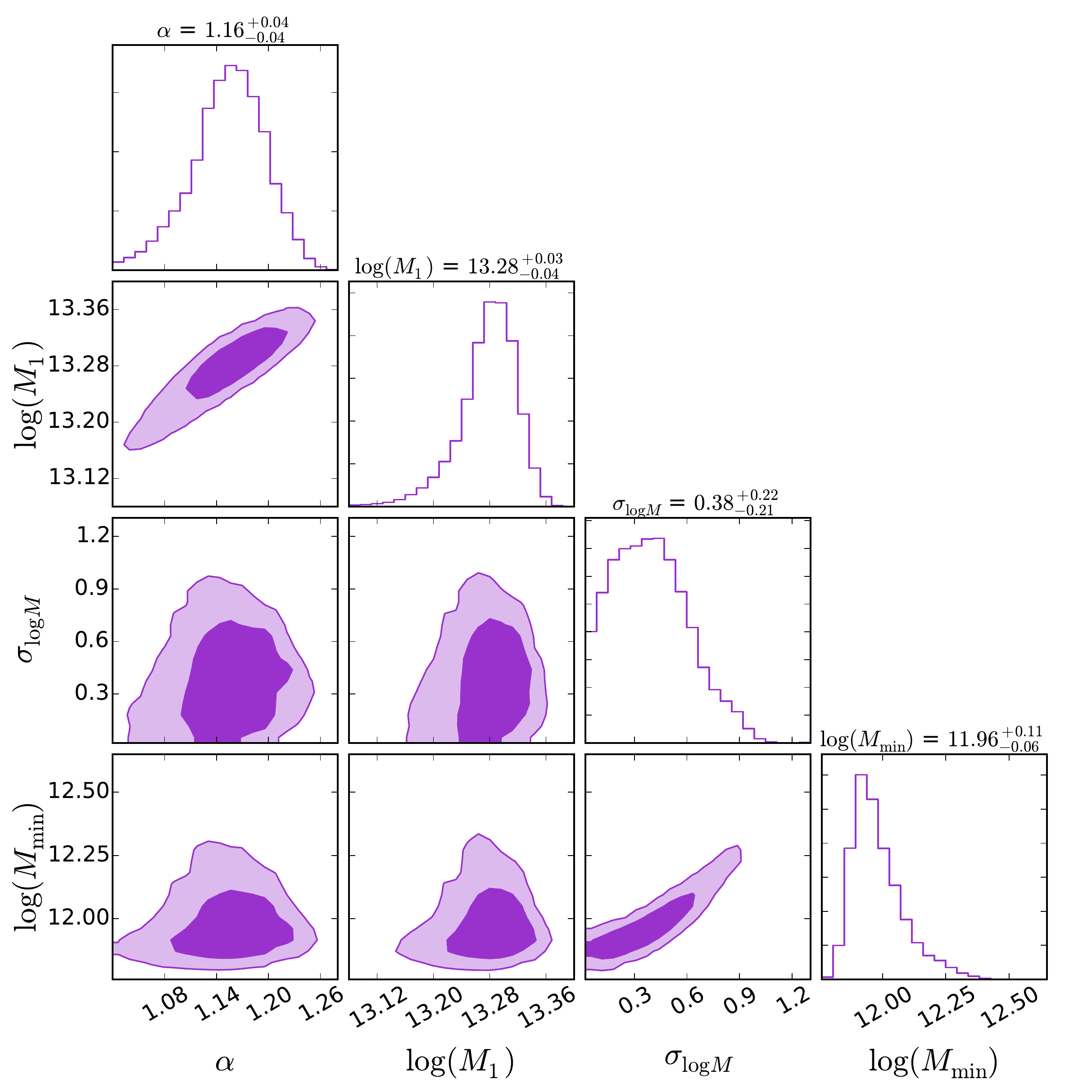}
\caption{
Two-dimensional marginalized constraints on HOD parameters inferred from
standard HOD fits to $\wprp(\rp)$ data for the $M_r<-20$ sample. The HOD parameter
$\log (M_0)$ is extremely poorly constrained by the $\wprp(\rp)$ data and has been
omitted. The inner contours contain $68\%$ of the posterior probability while the
outer contours contain $95\%$ of the probability. The panels along the diagonal
show the one-dimensional, marginalized posteriors on each of these parameters.
The values above each panel on the diagonal show the
median value for the parameter in our chains along with the
16$^{\rm th}$ and 84$^{\rm th}$ percentiles.
}
\label{fig:Mr20triangle}
\end{center}
\end{figure*}

\begin{figure*}
\begin{center}
\includegraphics[width=8.3cm]{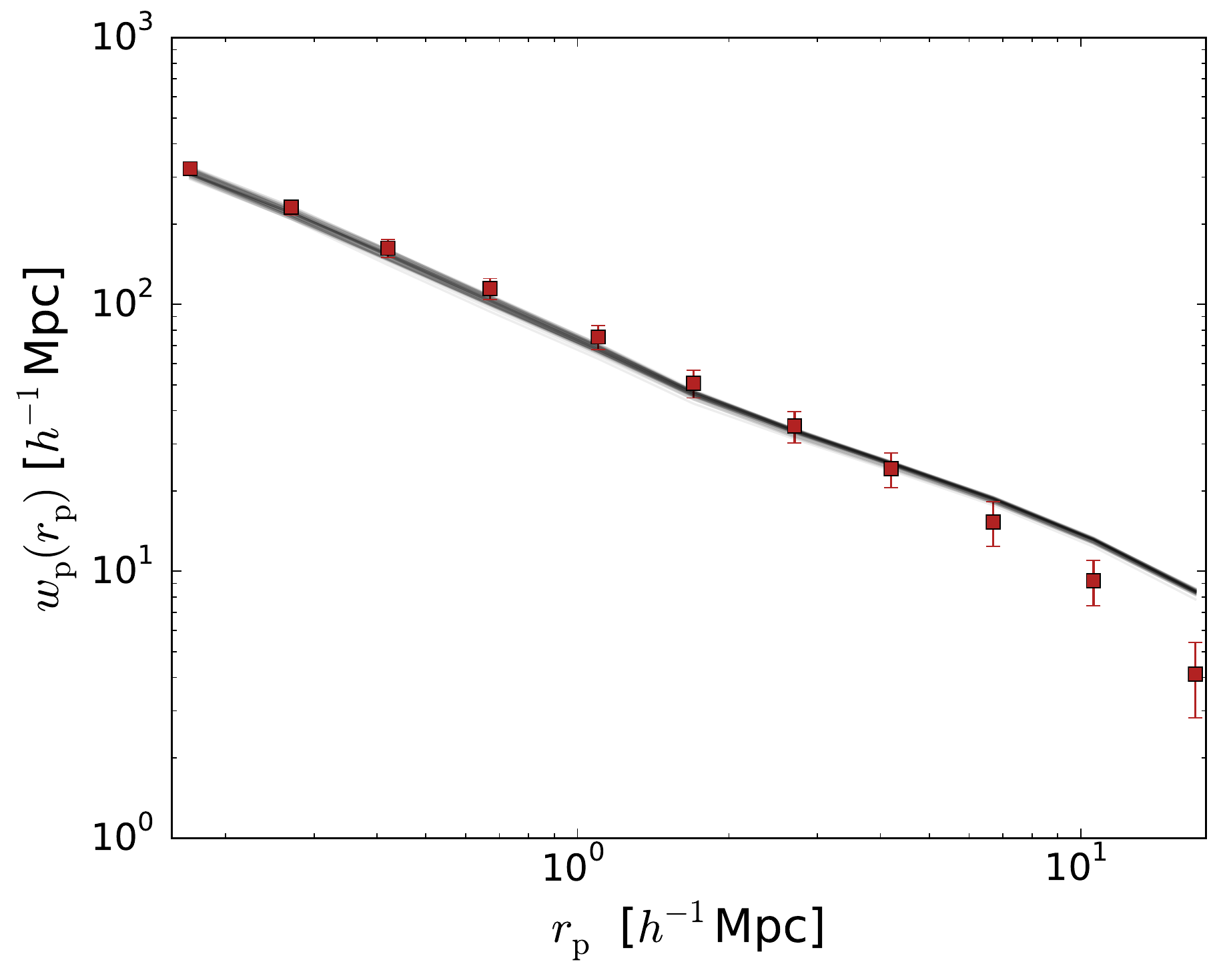}
\includegraphics[width=8.3cm]{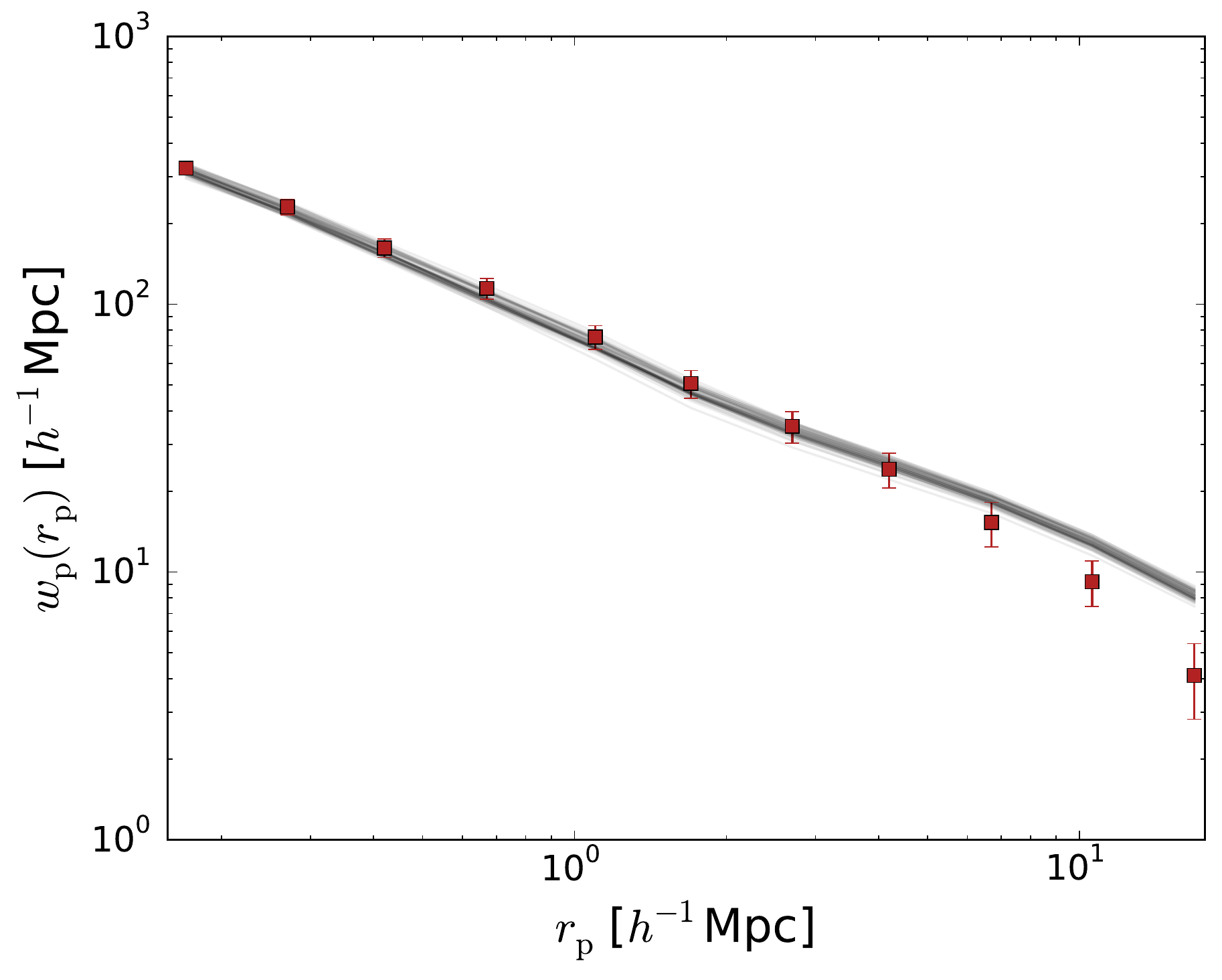}
\caption{
{\bf Left:} The $M_r<-19$ threshold sample projected correlation function with diagonal elements of
covariance (points with errorbars). The grey lines are 50 randomly-selected HOD models that yield
$\Delta \chi^2 < 5.89$ compared to the best-fitting model. For a Gaussian posterior distribution,
$\Delta \chi^2 = 5.89$ would contain 68\% of the probability in the full five-dimensional parameter
space. {\bf Right:} Same as the left panel but using a
fit to a Decorated HOD model that contain parameters 
to describe the strength of assembly bias.
}
\label{fig:Mr19samples}
\end{center}
\end{figure*}

\begin{figure*}
\begin{center}
\includegraphics[width=8.3cm]{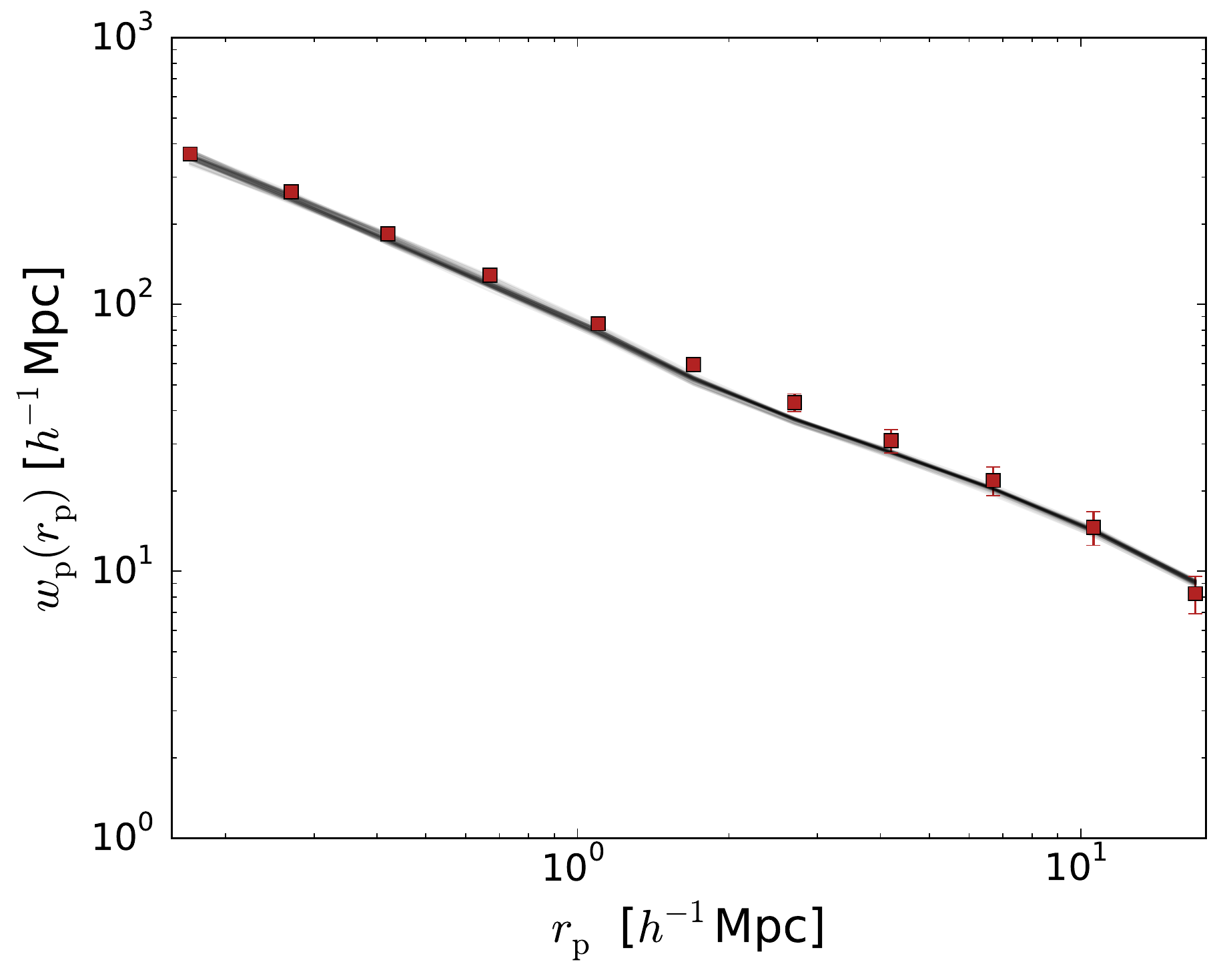}
\includegraphics[width=8.3cm]{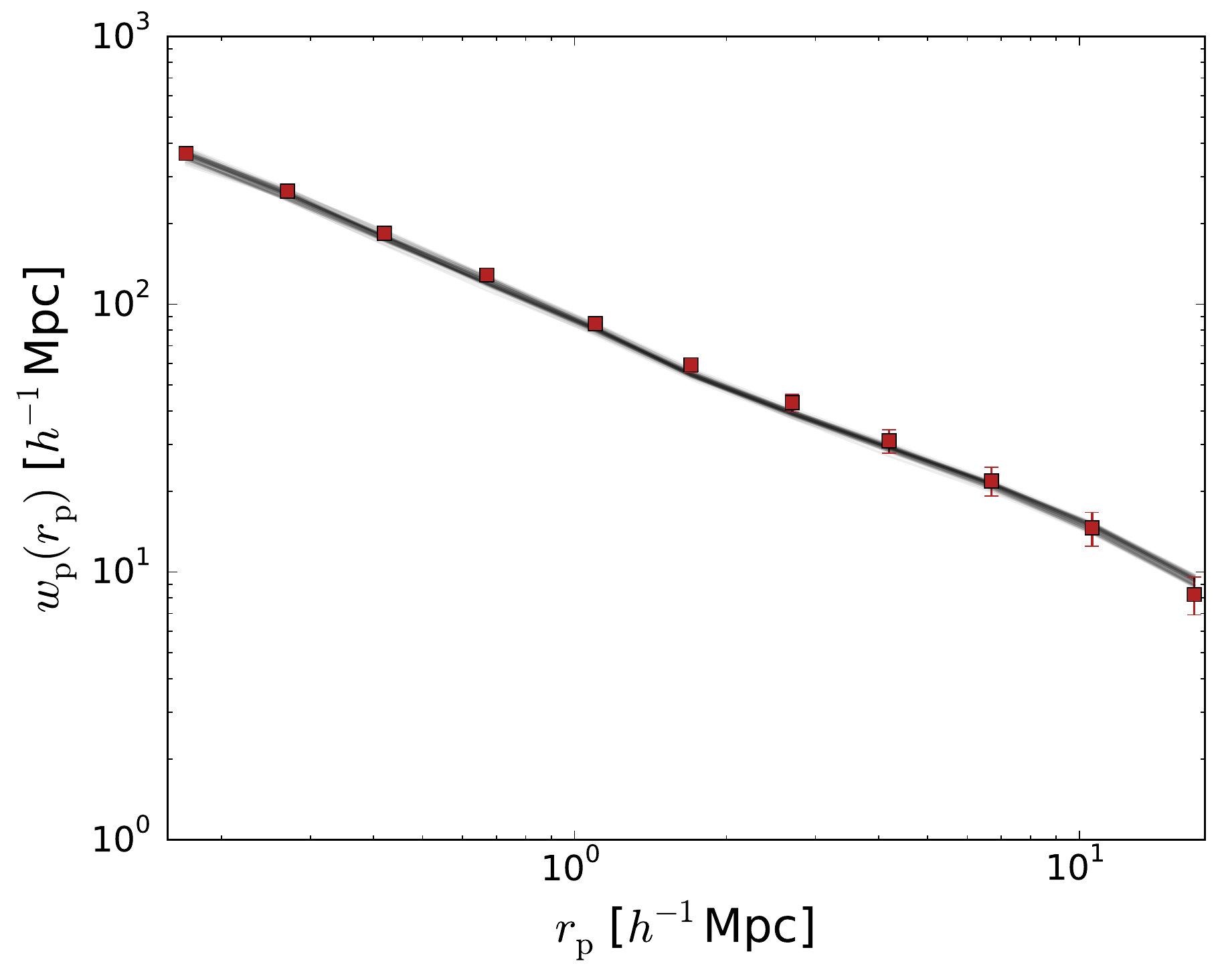}
\caption{
The same as Figure~\ref{fig:Mr19samples}, but for the $M_r<-20$ threshold sample.
}
\label{fig:Mr20samples}
\end{center}
\end{figure*}

\begin{figure*}
\begin{center}
\includegraphics[width=8.3cm]{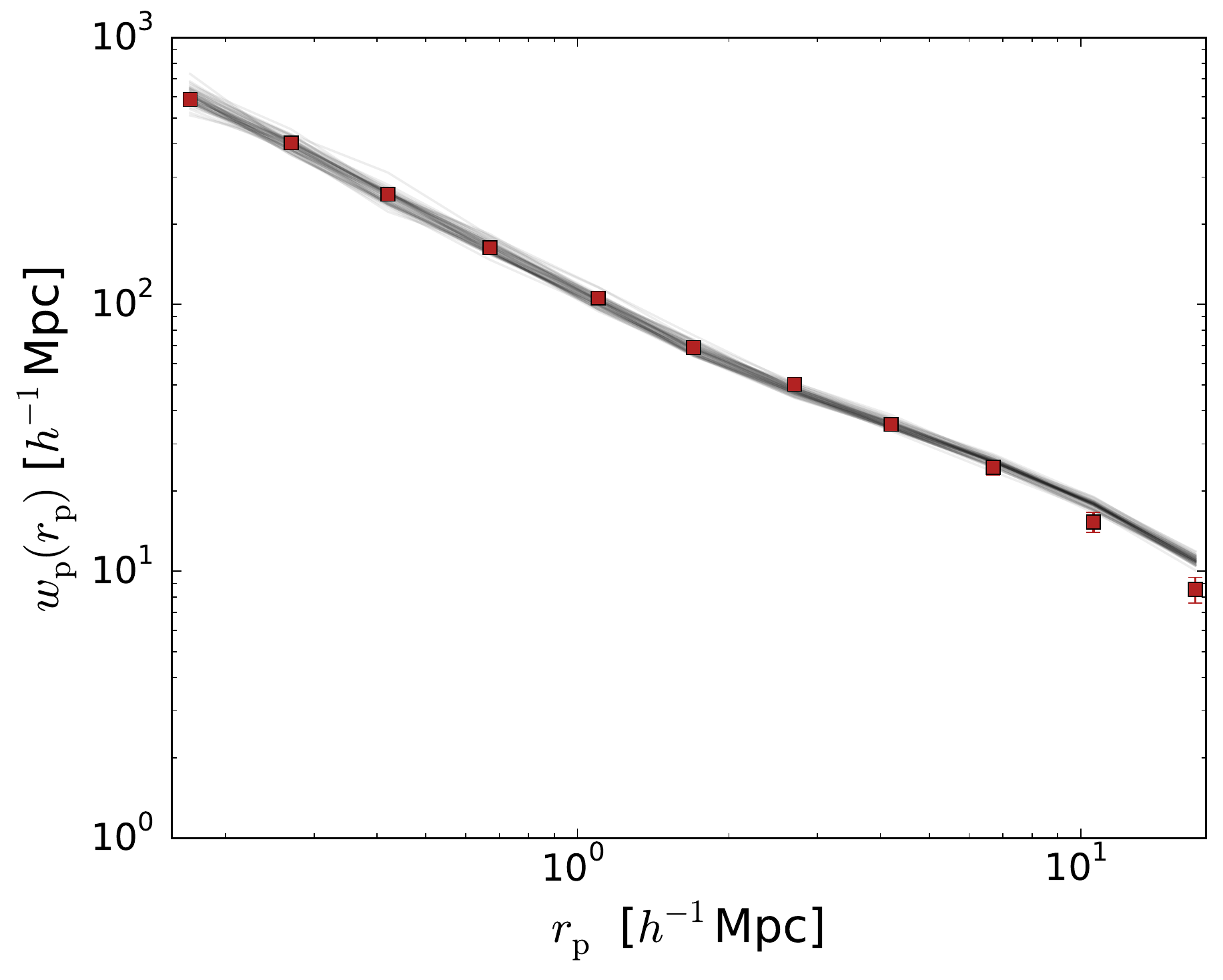}
\includegraphics[width=8.3cm]{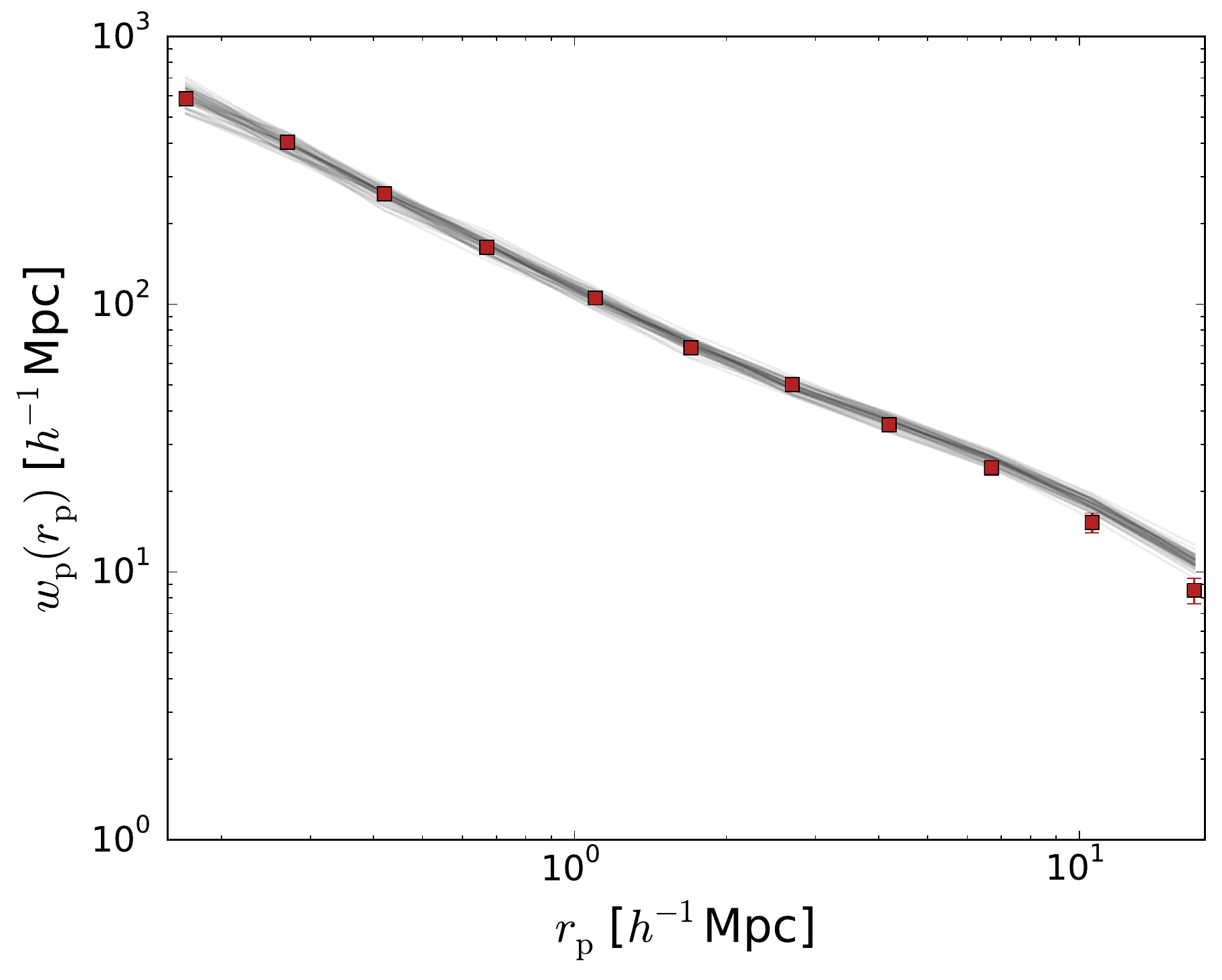}
\caption{
The same as Figure~\ref{fig:Mr19samples}, but for the $M_r<-21$ threshold sample.
}
\label{fig:Mr21samples}
\end{center}
\end{figure*}

Prior to discussing our results using models that include assembly bias, we present
results of standard HOD analyses that include no treatment of assembly bias. 
In our standard HOD analyses, the parameters $\log (M_{\rm min})$,
$\sigma_{\log M}$, $\alpha$, $\log (M_1)$, and $\log (M_0)$ are permitted to vary. 
The results of the standard HOD analyses and all other analyses are shown in the form of 
marginalized constraints on individual parameters in Table~2. Though the quality of 
our fits, as measured by the minimum of $\chi^2$ per Degree of Freedom (DoF), varies from case to case, 
all have a probability $\gtrsim 1\% $ of obtaining a higher value of $\chi^2$ by chance. 
An example of the inferred posteriors for the HOD parameters is shown in 
Figure~\ref{fig:Mr20triangle} for the sample with $\magr < -20$. Figure~\ref{fig:Mr20triangle} and 
all similar plots in this paper were made using a slightly modified version of the 
{\tt corner} software package \citep{corner}. We do 
not show the full posteriors for all five threshold samples in the interest of brevity.
The left-hand panels of Figure~\ref{fig:Mr19samples}, Figure~\ref{fig:Mr20samples}, and Figure~\ref{fig:Mr21samples}
show the projected correlation function data along with predictions for $\wprp(\rp)$ from 50 randomly-selected
models from the MCMC chains within $\Delta \chi^2 \le 5.89$ of the best-fit model. Under the assumption of 
a Gaussian posterior distribution, $\Delta \chi^2 = 5.89$ contains 68\% of the posterior probability for a 
five-parameter model. Note that the significant covariance in the data makes it difficult to
determine the quality of fits from visual inspection of these figures.

%--------------------------No AB Triangle Plot-----------------------------------------------------------------------------
%----------------------------------------------------------------------------------------------------------------------------------

Our parameter constraints can be compared to the standard HOD analysis performed by \citet{zehavi_etal11} 
by examining the top two rows in each luminosity threshold grouping in Table~2. In all 
cases, we quote the medians of our posteriors as our central values and our, 
generally asymmetric, errorbars, give the 16$^{\rm th}$ and 
84$^{\rm th}$ percentiles of the posterior samples so as to correspond roughly with ``$1-\sigma$" errors. 
The inferred parameters from our standard analyses differ in several ways from the \citet{zehavi_etal11}
analysis. Firstly, in our re-analysis of the projected clustering data, we generally find all mass scales to
be slightly higher than in the work of \citet{zehavi_etal11}. This difference is primarily due to the slightly
different cosmological model used in this work as compared to that of \citet{zehavi_etal11}.
The most important differences are in the values of $\Omega_{\rm M}$, and $\sigma_8$.
\citet{zehavi_etal11} assumed $\Omega_{\rm M}=0.25$ and $\sigma_8=0.8$,
whereas in the present work, we use the BolshoiP simulation in which $\Omega_{\rm M}=0.307$ and $\sigma_8=0.82$.
Slightly larger mass scales are necessary in an analysis with higher $\Omega_{\rm M}$ and $\sigma_8$ in order
to maintain galaxy number densities fixed with larger halo number densities. A detailed comparison between our central 
values and those of \citet{zehavi_etal11} is further confounded by the fact that we use a virial halo definition, 
whereas the analytic model of \citet{zehavi_etal11} 
has been calibrated to friend-of-friends halo masses (Z. Zheng, Private Communication).

%-------------------------------------------------------------------------------------------------------------------------------------------------
\begin{table*}
{\renewcommand{\arraystretch}{1.3}
\renewcommand{\tabcolsep}{0.2cm}
\begin{tabular}{l l c c c c c c c}
\hline
\hline
Sample $M_r$ &  Authors & $\log (M_{\rm min})$ & $\sigma_{\log M}$ & $ \log (M_1)$ & $\alpha$ & $A_{\rm cen}$ & $A_{\rm sat}$ & $\chi^2/\mathrm{DoF}$\\
\hline
$-21$ & Zehavi+11 & $12.78 \pm 0.10$ & $0.68 \pm 0.15$ & $13.80 \pm 0.03$ & $1.15 \pm 0.06$ & $--$ & $--$ & 3.1\\
$-21$ & Zentner+16 & $12.93^{+0.07}_{-0.10}$ & $0.74^{+0.09}_{-0.13}$ & $13.96^{+0.03}_{-0.05}$ & $1.27^{+0.08}_{-0.10}$ & $--$ & $--$ & 1.59\\
$-21$ & Zentner+16 & $12.83^{+0.11}_{-0.09}$ & $0.60^{+0.15}_{-0.17}$ & $13.93^{+0.05}_{-0.08}$ & $1.16^{+0.12}_{-0.14}$ & $0.29^{+0.44}_{-0.35}$ & $0.08^{+0.49}_{-0.36}$ & 1.34 \vspace*{6pt}\\
$-20.5$ & Zehavi+11 & $12.14 \pm 0.03$ & $0.17 \pm 0.15$ & $13.44 \pm 0.03$ & $1.15 \pm 0.03$ & $--$ & $--$ & 2.7\\
$-20.5$ & Zentner+16 & $12.25^{+0.07}_{-0.03}$ & $0.23^{+0.17}_{-0.15}$ & $13.59^{+0.02}_{-0.02}$ & $1.20^{+0.04}_{-0.04}$ & $--$ & $--$ & 1.90\\
$-20.5$ & Zentner+16 & $12.32^{+0.13}_{-0.08}$ & $0.45^{+0.21}_{-0.25}$ & $13.59^{+0.04}_{-0.04}$ & $1.14^{+0.05}_{-0.06}$ & $>0.08 (90\%)$ & $0.22^{+0.40}_{-0.31}$ & 1.40\\
             &                     &                                         &                                        &                                        &                                      & $>0\, (92.3\%)$ &                                           &
 \vspace*{8pt}\\
$-20$ & Zehavi+11 & $11.83 \pm 0.03$ & $0.25 \pm 0.11$ & $13.08 \pm 0.03$ & $1.00 \pm 0.05$ & $--$ & $--$ & 2.1\\
$-20$ & Zentner+16 & $11.96^{+0.11}_{-0.06}$ & $0.38^{+0.22}_{-0.21}$ & $13.28^{+0.03}_{-0.04}$ & $1.16^{+0.04}_{-0.04}$ & $--$ & $--$ & 1.88\\
$-20$ & Zentner+16 & $12.24^{+0.29}_{-0.21}$ & $0.84^{+0.37}_{-0.31}$ & $13.19^{+0.06}_{-0.08}$ & $1.05^{+0.06}_{-0.08}$ & $ >0.29 (99\%)$ & $0.01^{+0.35}_{-0.27}$ & 1.09\\
          &                     &                                         &                                       &                                         &                                      & $>0\, (99.9\%)$     &                                      & \vspace*{8pt}\\
%$-20^d$ & Zentner+16 & $11.91^{+0.10}_{-0.03}$ & $0.26^{+0.25}_{-0.17}$ & $13.34^{+0.04}_{-0.05}$ & $1.25^{+0.05}_{-0.06}$ & $--$ & $--$ & 0.70 \\
%$-20^d$ & Zentner+16 & $12.00^{+0.29}_{-0.12}$ & $0.51^{+0.42}_{-0.32}$ & $13.31^{+0.06}_{-0.09}$ & $1.17^{+0.08}_{-0.09}$ & $0.76^{+0.18}_{-0.43}$ & $0.11^{+0.37}_{-0.34}$ & 0.30\vspace*{3pt}\\
$-19.5$ & Zehavi+11 & $11.57 \pm 0.04$ & $0.17 \pm 0.13$ & $12.87 \pm 0.03$ & $0.99 \pm 0.04$ & $--$ & $--$ & 1.00 \\
$-19.5$ & Zentner+16 & $11.76^{+0.33}_{-0.11}$ & $0.51^{+0.51}_{-0.29}$ & $13.05^{+0.04}_{-0.08}$ & $1.12^{+0.04}_{-0.07}$ & $--$ & $--$ & 1.24\\
$-19.5$ & Zentner+16 & $11.80^{+0.36}_{-0.16}$ & $0.63^{+0.53}_{-0.37}$ & $13.04^{+0.09}_{-0.12}$ & $1.06^{+0.07}_{-0.10}$ & $>-0.01 (84\%)$ & $>-0.16 (84\%)$ & 0.69 \vspace*{6pt}\\
$-19$ & Zehavi+11 & $11.45 \pm 0.04$ & $0.19 \pm 0.13$ & $12.64 \pm 0.04$ & $1.02 \pm 0.02$ & $--$ & $--$ & 1.8 \\
$-19$ & Zentner+16 & $11.72^{+0.33}_{-0.19}$ & $0.70^{+0.51}_{-0.45}$ & $12.78^{+0.04}_{-0.04}$ & $1.03^{+0.04}_{-0.04}$ & $--$ & $--$ & 2.67\\
$-19$ & Zentner+16 & $11.62^{+0.33}_{-0.13}$ & $0.53^{+0.57}_{-0.35}$ & $12.83^{+0.06}_{-0.07}$ & $1.02^{+0.04}_{-0.04}$ & $0.35^{+0.45}_{-0.66}$ & $>0.02 (84\%)$ & 2.01\\
          &                      &                                       &                                        &                                         &                                      &                                       & $>0\, (85\%)$ & \\
%$-19^d$ & Zentner+16 & $11.87^{+0.17}_{-0.27}$ & $1.10^{+0.30}_{-0.52}$ & $12.84^{+0.17}_{-0.23}$ & $1.11^{+0.15}_{-0.15}$ & $--$ & $--$ & 2.30 \\
%$-19^d$ & Zentner+16 & $11.75^{+0.24}_{-0.27}$ & $0.94^{+0.41}_{-0.55}$ & $12.99^{+0.14}_{-0.18}$ & $1.17^{+0.14}_{-0.14}$ & $-0.17^{+0.50}_{-0.47}$ & $$ & 1.70\\
\hline
\end{tabular}
\medskip
\caption{
Results of standard HOD fits to SDSS DR7 $w_{\rm p}(r_{\rm p})$ as well as
fits using a parameterized model of assembly bias.
Assembly bias is quantified by the parameters $A_{\rm cen}$ ($A_{\rm sat}$) for central (satellite) galaxies. The secondary
property that we assume to determine the galaxy HOD is halo concentration. $A_{\rm cen,sat}=0$ means that there is no
assembly bias while $A_{\rm cen,sat}=1$ ($A_{\rm cen,sat}=-1$) means that galaxy abundance is maximally
correlated (anticorrelated) with halo concentration at fixed $\mvir.$
Thus the $A_{\rm cen,sat}$ parameters span the range $[-1, 1].$
If the constraints on $A_{\rm cen}$ and $A_{\rm sat}$ are unspecified in the table, 
then the model used to interpret the data
does not include assembly bias. In our analyses, quoted parameter values with errors correspond to the median
value of the parameter and the 16$^{\rm th}$ and 84$^{\rm th}$ percentiles. In cases for which the posterior
on $A_{\rm cen,sat}$ is significant at the boundary of the permissible parameter range (e.g., $A_{\rm cen}$ for $\magr < -20$), 
we provide one-sided constraints. In the cases exhibiting the strongest assembly bias, we also quote the probability with 
which the inferred value of $A_{\rm cen,sat}$ exceeds zero.
}
}
 \label{table:parameters}
\end{table*}
%---------------------------------------------------------------------------------------------------------------------------------

%----------------------------------------------------------------------------------------------------------------------------------
%---------------------------------------------------------------------------------------------------------------------------------

%---------------------------------------------------------------------------------------------------------------------------------
%---------------------------------------------------------------------------------------------------------------------------------

%---------------------------------------------------------------------------------------------------------------------------------
%---------------------------------------------------------------------------------------------------------------------------------

A second noteworthy difference between the present work and that of
\citet{zehavi_etal11} is that we find many parameters to be notably
more poorly constrained. At the lower luminosity thresholds, for
example, we constrain $\log (M_{\rm min})$ and $\sigma_{\log M}$ with
several times lower precision than \citet{zehavi_etal11}. We do not
show our constraints on $\log (M_0)$ as they are very poor, with
1-sigma constraints $\gta 1$~dex for all samples. In several cases,
the constraint on $\log (M_0)$ is determined by the prior given in
Table~1. This is in stark contrast to several of the
results of \citet{zehavi_etal11}. For example, for the threshold
sample with $M_r < -19.5$ ($M_r < -20.5$), \citet{zehavi_etal11} quote
$\log (M_0) = 12.23 \pm 0.17$ ($12.35 \pm 0.24$), whereas we infer
$\log (M_0) = 11.38^{+0.95}_{-1.57}$
($11.19^{+0.89}_{-1.39}$). Examining the form of Eq.~(\ref{eq:nsat}),
it is sensible that the parameter $\log (M_0)$ should be unconstrained
at the lower end, because the value of $M_0$ does not alter the
predicted satellite number once $M_0 \ll M_1$. The tighter constaints are also 
particularly puzzling given that \citet{zehavi_etal11} adopt an arbitrary 5\% error on 
all galaxy number density measurements, an error that greatly exceeds the 
Poisson error, typically $< 1\%$, that we adopt. This additional error 
contribution also limits the value of comparing our $\chi^2$ values with those of 
\citet{zehavi_etal11}, as the latter values will be artificially lower. 

We have confirmed with one of the authors of \citet{zehavi_etal11}
that the MCMC chains used in a number of their analyses may not have been
properly converged and that this may have led to a significant underestimation of the
uncertainties on the inferred parameters, especially $\log (M_0)$, $\log (M_{\rm min})$,
and $\sigma_{\log M}$. The analysis of \citet{zehavi_etal11}
used $10^4$ samples, whereas we find that several $\times 10^5$ to $10^6$ 
samples are often necessary for convergence. In all cases, our final results are from 
$>10^6$ samples of the posterior. The cause of the problem is likely a 
restrictive proposal distribution that causes the chain to
diffuse through the posterior distribution only extremely slowly (Z. Zheng, Private Communication). 
We have recreated qualitatively similar behavior considering only small subsets of our full 
MCMC chains. Consequently, insufficient sampling of the posterior seems to be the likely resolution
of the discrepancies between our work and that of \citet{zehavi_etal11}.

Two degeneracies are manifest in Fig.~\ref{fig:Mr20triangle} that are common to all of our
analyses. The parameters $\log (M_1)$ and $\alpha$ are degenerate with each other and positively
correlated. The parameter $M_1$ is the mass scale at which a halo has one satellite on
average, and $\alpha$ is the power-law index describing the dependence of average satellite
number on halo mass. Increasing $M_1$ {\em decreases} the number of satellites in massive halos
by increasing the mass scale where the power law abundance becomes operative. 
An increase in $\alpha$ can partly
compensate for an increase in $M_1$ by increasing 
the rate at which average satellite number
grows with halo mass.

Look for the purple banana\footnote{'Til they put us in the truck.} in the third panel from the left in 
the bottom row of Figure~\ref{fig:Mr20triangle}. This panel illustrates that the parameters
$\log (M_{\rm min})$ and $\sigma_{\log M}$ share a relatively narrow degeneracy as well.
This degeneracy is largely induced by the measured
number density of the sample. Increasing $\log (M_{\rm min})$ decreases galaxy
number density, but this can be compensated by an increase in $\sigma_{\log M}$, which
places galaxies in a fraction of the considerably more numerous halos with masses less
than $M_{\rm min}$. The consequence is that $\log (M_{\rm min})$ and $\sigma_{\log M}$ are
degenerate with each other such that most of the posterior probability lies in a narrow band
along which $\log (M_{\rm min})$ and $\sigma_{\log M}$ are positively correlated.
In the following plots, we exclude the parameter $\sigma_{\log M}$,
in order to increase the clarity of the plots, because the viable range of $\sigma_{\log M}$ is
determined by this simple degeneracy with $\log (M_{\rm min})$. 
Our constraints on $\sigma_{\log M}$ are listed in Table~2.

The results of this subsection demonstrate that we achieve reasonable fits to projected galaxy clustering
data using direct HOD population of a high-resolution numerical simulation of structure formation. These results
also update and supersede existing constraints in the literature in at least three respects. First, we work within the
best-fit Planck cosmology. Second, we perform our parameter inference analysis using direct population of halos
identified in a numerical simulation of cosmological structure formation (BolshoiP). This greatly mitigates modeling
uncertainties associated with nonlinear density field evolution, scale-dependent halo bias, halo exclusion,
or other effects that have been difficult to incorporate into analytical halo models with high precision
\citep[e.g.,][and references the]{vdBosch13}. Third,
we have explored the posteriors of the parameters with significantly more samples (roughly two orders of magnitude),
thereby mitigating errors on inferred parameters caused by insufficient sampling of the
posterior by \citet{zehavi_etal11}.

%-------------------------
\subsection{Analysis with Decorated HOD}
\label{subsection:ab}
%-------------------------

We turn now to a discussion of our parameter inference analysis of projected galaxy clustering
in decorated HOD models that include a treatment of galaxy assembly bias. In this work,
we consider only the simplest model of galaxy assembly bias, introducing only two new
parameters, $A_{\rm cen}$ and $A_{\rm sat}$, that describe the strength of central galaxy
and satellite galaxy assembly bias respectively. These parameters are limited to values
of $-1 \le A_{\rm cen,sat} \le 1$, and $A_{\rm cen,sat}=0$ when there is no galaxy assembly bias.
In this work, we use halo concentration as our secondary halo property, so $A_{\rm cen,sat}=1$
($A_{\rm cen,sat}=-1$) means that the mean number of galaxies per halo is maximally
correlated (anti-correlated) with halo concentration. The model and its implementation
in {\tt Halotools} is discussed further in Section~\ref{subsection:decorated} above 
and in more detail in \citet{hearin_etal16}.

Examples of our fits are given in the right-hand panels of Figure~\ref{fig:Mr19samples},
Figure~\ref{fig:Mr20samples}, and Figure~\ref{fig:Mr21samples}. The general trend that
can be gleaned from these figures is that introducing assembly bias improves the fit to
the measured two-point functions across the
transition from the one-halo (highly nonlinear) to two-halo (nearly linear) regimes near 
$\rp \sim 2\, h^{-1}{\mathrm{Mpc}}$, an effect that could be anticipated by the known 
scale-dependence of galaxy assembly bias \citep{sunayama_etal16}. 
This is most apparent for the $M_r < -20$ threshold sample shown in Fig.~\ref{fig:Mr20samples}.
Visually, these differences appear to be small; however, Table~2 shows that
they are statistically important.

\begin{figure*}
\begin{center}
\includegraphics[width=15.0cm]{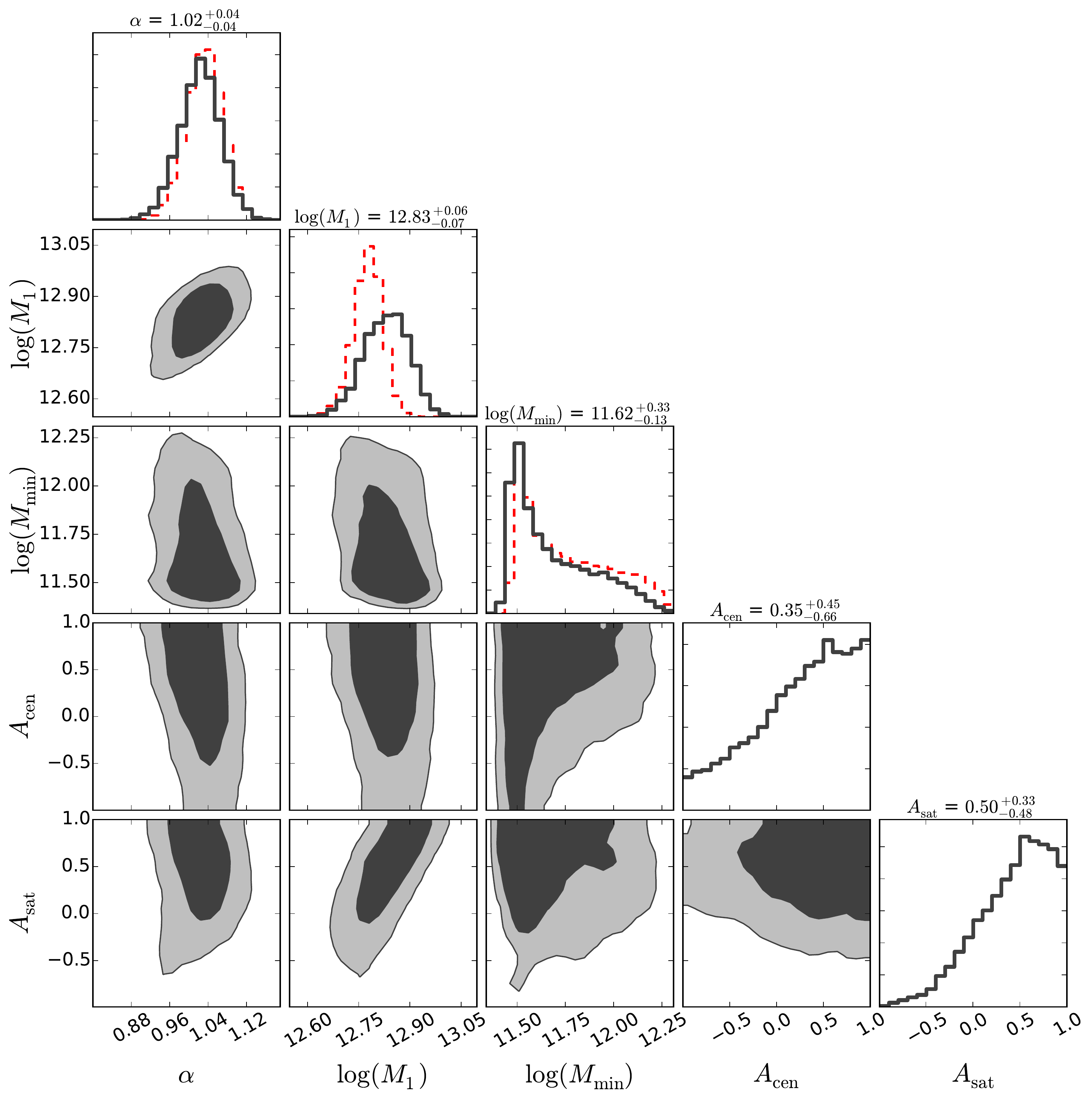}
\caption{ Two-dimensional marginalized constraints on decorated HOD
  parameters inferred from fits to $\wprp(\rp)$ data for the $M_r<-19$
  sample. The contours and histograms along the diagonal panels are as
  in Fig.~\ref{fig:Mr20triangle}. The dashed line in the panels along the diagonal
  show the posteriors on these parameters from the standard analysis without
  any parameterization of galaxy assembly bias. In these panels, both histograms 
  are normalized to the same total probability. The decorated HOD models include a
  two-parameter model for assembly bias. The HOD parameter 
  $\log (M_0)$ is extremely poorly constrained by the data and we omit it for 
  clarity. Likewise, as in Fig.~\ref{fig:Mr20triangle},
  $\sigma_{\log M}$ and $\log (M_{\rm min})$ share a narrow
  degeneracy, so we have suppressed $\sigma_{\log M}$ in order to make
  constraints on other parameters more easily visible.  }
\label{fig:Mr19ABtriangle}
\end{center}
\end{figure*}

\begin{figure*}
\begin{center}
\includegraphics[width=15.0cm]{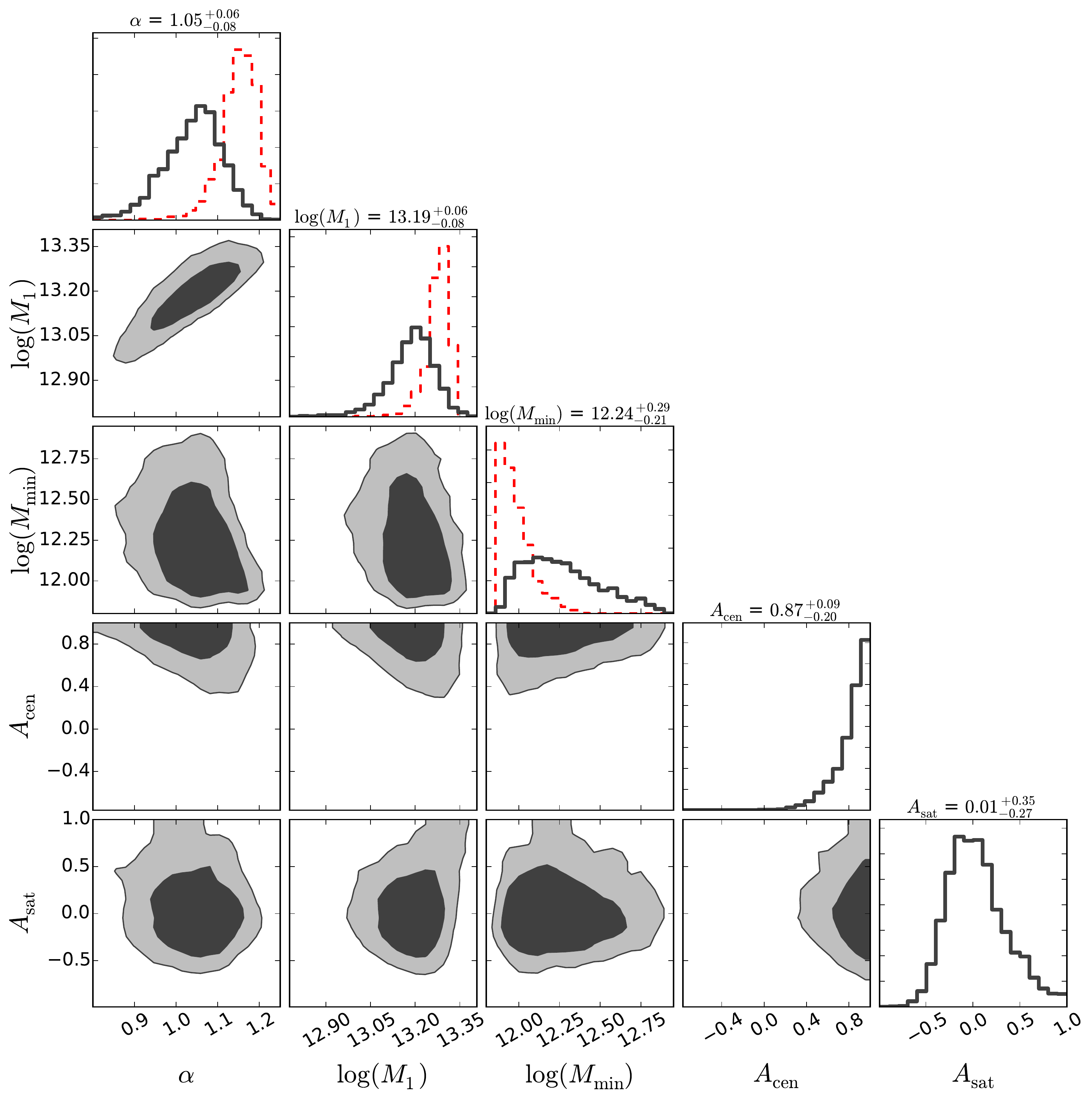}
\caption{
The same as Figure~\ref{fig:Mr19ABtriangle}, but for the $M_r<-20$ sample.
}
\label{fig:Mr20ABtriangle}
\end{center}
\end{figure*}

\begin{figure*}
\begin{center}
\includegraphics[width=15.0cm]{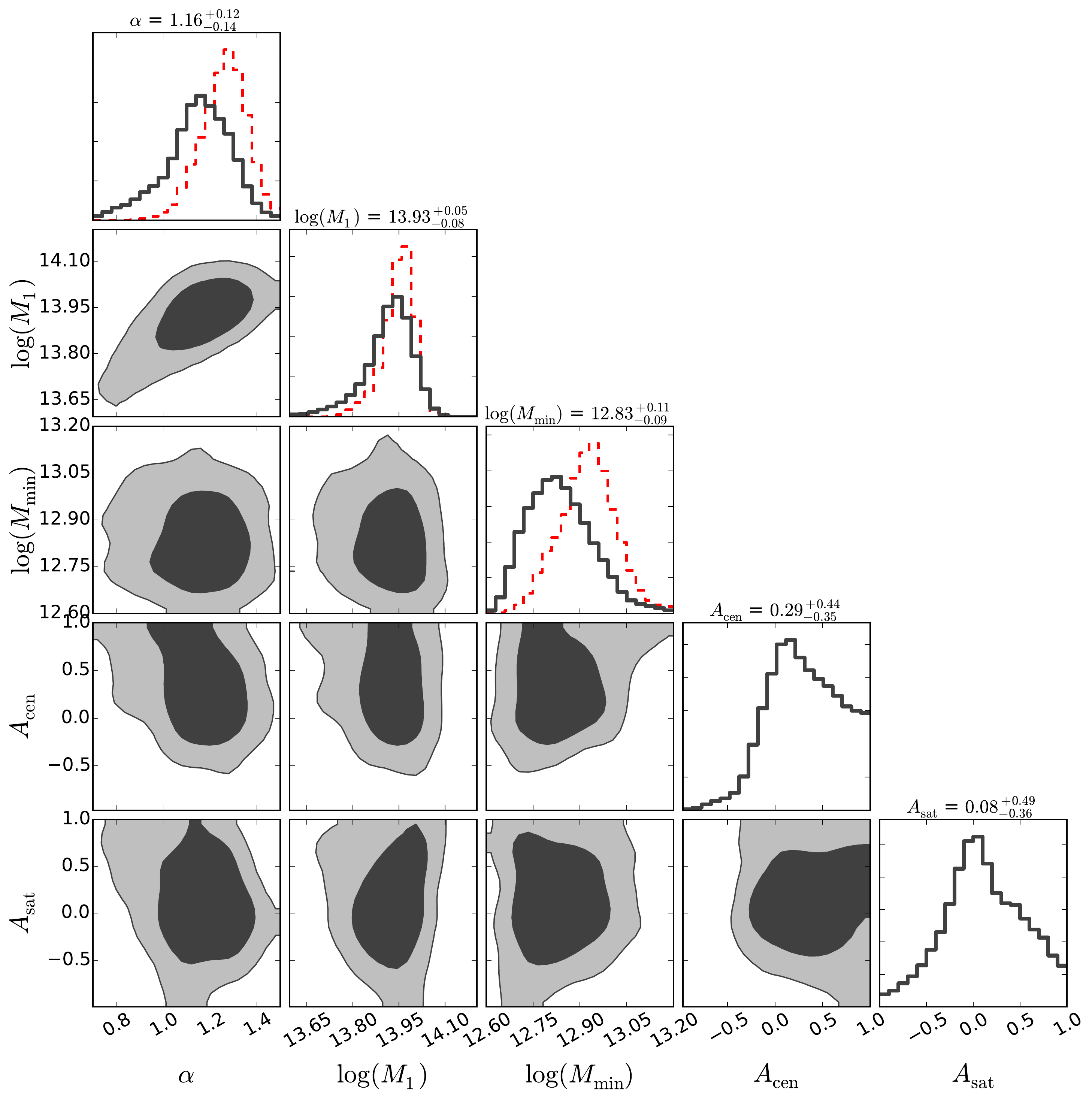}
\caption{
The same as Figure~\ref{fig:Mr19ABtriangle}, but for the $M_r<-21$ sample.
}
\label{fig:Mr21ABtriangle}
\end{center}
\end{figure*}

The one-dimensional marginalized constraints on all parameters from
these analyses are given in the lowest row of each luminosity
threshold grouping in Table~2.  In cases where
the posterior distribution of either $A_{\rm cen}$ or $A_{\rm sat}$ contains significant 
probability at the limits of the parameter range (e.g., see the posterior distribution for
$A_{\rm cen}$ in Fig.~\ref{fig:Mr20ABtriangle}), we quote one-sided 
constraints.  Examples of one- and
two-dimensional visualizations of the posteriors from our analyses are
shown in Figure~\ref{fig:Mr19ABtriangle},
Figure~\ref{fig:Mr20ABtriangle}, and Figure~\ref{fig:Mr21ABtriangle}
for the $M_r<-19$, $M_r<-20$, and $M_r<-21$ samples respectively.

%------------------------------------------------------------------------------------------------
%----------------------------------------------------------------------------------------------

%---------------------------------------------------------------------------------------------------
%---------------------------------------------------------------------------------------------------

%=-------------------------------------------------------------------------------------------------
%----------------------------------------------------------------------------------------------------

Table~2 and Figures~\ref{fig:Mr19ABtriangle}--\ref{fig:Mr21ABtriangle}
all make several simple, generic points. Introducing additional parameter
freedom associated with galaxy assembly bias generally increases the volume of the 
viable parameter space, even for the subset of five standard HOD parameters. 
Constraints on the standard HOD parameters are generally less restrictive in models that 
include assembly bias. This is exactly what is expected from the introduction of additional 
parameter freedom.

%=-------------------------------------------------------------------------------------------------
\begin{figure*}
\begin{center}
\includegraphics[width=8.0cm]{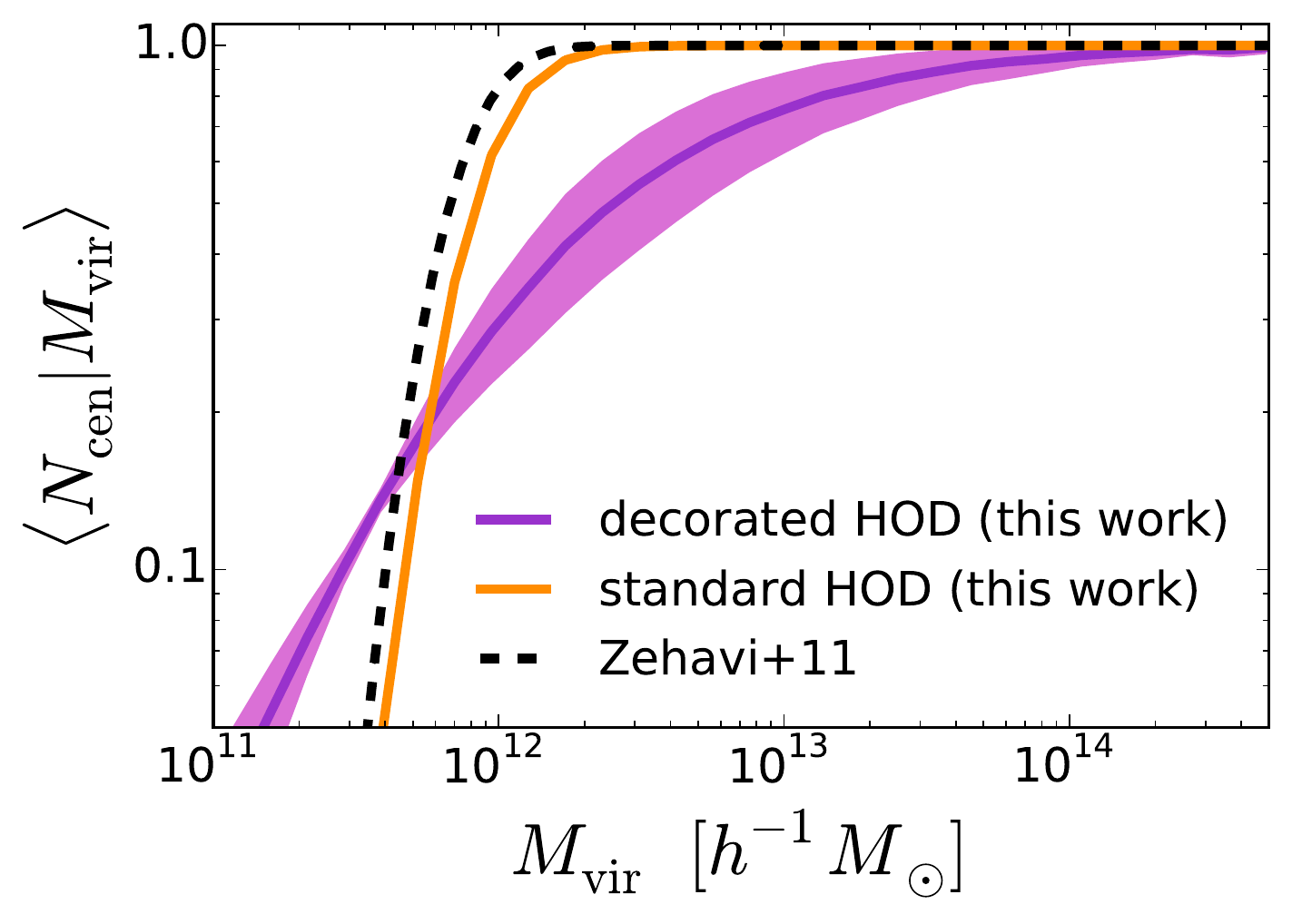}
\includegraphics[width=8.5cm]{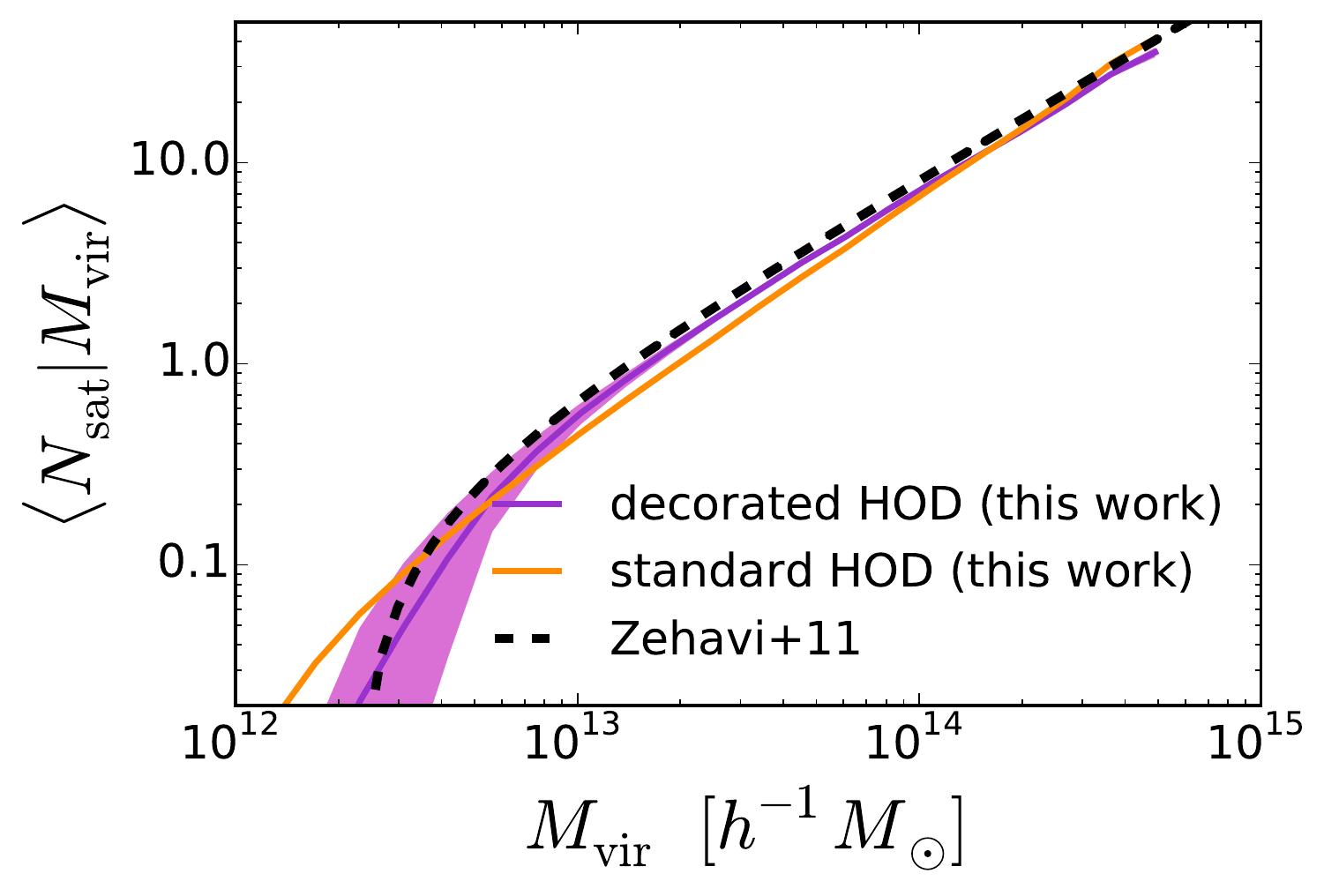}
\caption{
The first moments of the central (left panel) and satellite (right) 
occupation distributions from the best-fit models to the 
$\magr \le -20$ sample. The solid, purple curve shows the HOD of our best-fit model
that includes the effect of assembly bias; the solid, orange curve shows the HOD
of the best-fit model in which assembly bias is assumed to be zero;
the dashed, black curve shows the best-fit HODs from \citet{zehavi_etal11}.
The light purple error band around the decorated HOD results contains 68\% of 
all samples of the posterior within $\Delta \chi^2 \le 5.89$ of the best-fit model at 
each bin of halo mass. For any individual model, 
the values of the HOD at each mass point are obviously correlated and do 
not vary independently from one mass bin to another. 
}
\label{fig:directhodcomparison}
\end{center}
\end{figure*}
%----------------------------------------------------------------------------------------------------

Focusing attention on the parameters describing galaxy assembly bias, it is
evident that these parameters are often quite poorly constrained by galaxy
clustering data. This is important as it implies that galaxy clustering of the precision
of SDSS DR7 measurements cannot rule out, or strongly restrict, galaxy assembly
bias in many cases. This is the case, for example, for the $M_r<-21$ sample with posteriors
shown in Fig.~\ref{fig:Mr21ABtriangle}. In this case, both $A_{\rm cen}$ and
$A_{\rm sat}$ have posteriors that peak near zero, but are very broad.
Nonetheless, it is apparent that the presence of galaxy assembly bias
can alter the inferred HOD, or more generally, the inferred relationship between
galaxies and halo mass. This is most evident for the $M_r < -20$ threshold sample,
for which there are significant differences in the inferred values of all baseline HOD
parameters between models with and without galaxy assembly bias. Other threshold
samples exhibit significant differences particularly for $\alpha$, and to a lesser degree
for $\log (M_{\rm min})$ and $\sigma_{\log M}$.

Beyond those generic conclusions, a few specific cases are worthy of further examination.
Consider the $M_r < -20$ sample. The inferred value of $A_{\rm cen} > 0.28$ at 99\%
confidence. In this case the data strongly prefer $A_{\rm cen} > 0$ ($A_{\rm cen} > 0$ at $99.9\%$) 
and thus strongly prefer galaxies to reside in halos of larger concentration at fixed halo mass. 
Figure~\ref{fig:directhodcomparison}
compares the inferred HODs for the $M_r<-20$ sample in a standard HOD analysis with the
inferred HOD from our decorated HOD analysis that includes parameters to describe
assembly bias. The bands in purple rein in the region of HODs for models within 
$\Delta \chi^2 \le 5.89$ of the best-fit model. 
This is an explicit example of the degree to which the inferred relationship
between galaxies and halos can be altered by assembly bias. In the assembly-biased case, the efficiency 
of forming a sufficiently luminous galaxy is a function that increases slowly over nearly two 
decades in halo mass in contrast to a standard analysis HOD where this efficiency rises 
rapidly over roughly one third of a decade in $\mvir$. Taking these results at face value, 
the reason that the HOD of central galaxies can vary slowly over several orders of magnitude 
in $\mvir$ is that larger galaxies form more efficiently in high-concentration halos at fixed 
mass.

Though the evidence for assembly bias is the most significant for the $M_r<-20$ sample,
there are hints of galaxy assembly bias in other samples. Satellite galaxies show a marginal preference
for occupying halos of higher concentration in the $M_r < -19$ threshold sample, with $A_{\rm sat}>0$ at 
85\%. The $M_r < -19.5$ sample exhibits weak preference for a positive correlation of galaxy occupation 
with halo concentration at fixed mass for both satellite galaxies and central galaxies. The low value of 
$\chi^2$ in this case is also suggestive of possible over-fitting using these additional parameters (but see 
the discussion in the subsequent paragraph). Continuing upward with luminosity,
the $M_r < -20.5$ sample exhibits a significant preference for central galaxy assembly bias, 
with $A_{\rm cen}>0$ at 92.3\%. Lastly, there is no preference for either central galaxy or 
satellite galaxy assembly bias for the $M_r < -21$ threshold sample, 
for which both $A_{\mathrm{cen}}$ and $A_{\mathrm{sat}}$
are consistent with zero well within 1$\sigma$. While we give a brief discussion of these results 
in the following section, the implications of all of these results, particularly for 
quantities derived from HODs, are far too numerous 
to address comprehensively in the present manuscript; 
however, we plan to address them in forthcoming publications.

%-------------------------------------------------------------------------------------------------------------------------------------------------
\begin{table}
\begin{center}
{\renewcommand{\arraystretch}{1.3}
\renewcommand{\tabcolsep}{0.2cm}
\begin{tabular}{c c}
\hline
\hline
$M_r$ Threshold & $\Delta \mathrm{BIC}$ \\
\hline
-21 & -0.54 \\
-20.5 & 1.33 \\
-20 & 2.88\\
-19.5 & 0.26\\
-19 & 3.78\\
\hline
\end{tabular}
\medskip
\caption{
Change to the Bayesian Information Criterion, $\Delta \mathrm{BIC}$,
after introducing additional parametric freedom to accommodate galaxy
assembly bias. Sign convention is such that positive values favor models including
assembly bias, negative values favor standard HOD models with no assembly bias parameters. Changes in
the Bayesian Information Criterion $\vert \Delta \mathrm{BIC}\vert \ge 5$ strongly favor one model
over another. None of the models we explore achieve this threshold.
}
 }
 \label{table:bic}
 \end{center}
\end{table}
%--------------------------------------------------------------------------------------------------------------------------------

In addition to fitting model parameters, it is possible to ask whether or not the fits to the data warrant the use of 
additional parameters (in this case, the two parameters used to describe assembly bias). This requires a 
metric that penalizes the use of additional parameters. A commonly-used metric is the 
Bayesian Information Criterion 
\citep[BIC,][]{schwarz78}, defined as 
\begin{equation}
\mathrm{BIC} = -2 \mathcal{L}_{\rm max} + k\, \ln N, 
\end{equation}
where $\mathcal{L}_{\rm max}$ is the maximum of the likelihood, 
$k$ is the number of parameters in the model and $N$ is the number of 
data points. We compute the change in the BIC, $\Delta$BIC, by subtracting 
the BIC for the analysis with assembly bias from the BIC in the standard analysis 
in which there is no assembly bias. With this convention, $\Delta \mathrm{BIC} > 0$ 
favors assembly bias, suggesting that the data warrant the additional parametric freedom. 
Conversely, $\Delta \mathrm{BIC} < 0$ favors the standard analysis, 
suggesting that the additional freedom in the model does not improve the description of the data 
to a sufficiently great degree to support the added complexity. 
A common convention is to refer to the preference as 
{\em strong} when $\vert \Delta \mathrm{BIC}\vert \ge 5$. 

The values of $\Delta$BIC are given in Table~3. In the analysis of the $\magr < -21$ sample, 
$\Delta \mathrm{BIC} < 0$, indicating that the additional parameters describing assembly bias are not 
warranted by the data. For all other samples, $\Delta \mathrm{BIC} > 0$, showing a preference for 
the models with assembly bias. The cases that most support the additional complexity 
of the decorated HOD are $\magr < -19$ and $\magr < -20$, for which $\Delta$BIC is 
considerable, but in no case does the $\Delta$BIC rise to the conventional level for {\em strong} 
preference.

%------------------------------------------------
\section{Interpretation}
\label{section:discussion}
%------------------------------------------------

The results of the previous section suggest that halo
mass is insufficient to predict the galaxy content of a halo to the
precision mandated by an analysis of SDSS DR7 clustering measurements
(particularly the $\magr < -19$, $\magr < -20$, and $\magr < -20.5$
samples).  This should not be surprising, as the process of galaxy
formation and evolution is complex and there is little reason to
suspect that all potentially-important factors are determined by a
single halo property. Our results further suggest that halo
concentration (or some other halo property correlated with halo
concentration) impacts the galaxy content of a halo, at least for
galaxies of some luminosities. In particular, the assembly bias
inferred from the $\magr < -20$ and $\magr < -20.5$ samples suggests
that galaxies in roughly Milky Way-sized halos tend to be more
luminous if they reside in halos with higher-than-average
concentrations.

Within the context of the standard model of structure formation, this
can be interpreted in at least two ways that are not mutually
exclusive. Although our analysis cannot support either of these
scenarios unambiguously, they are reasonable starting points for
further consideration.  One possibility is simply that the growth and
structure of the host halo itself may drive this
luminosity-concentration dependence. For example, halos of higher
concentration at fixed mass have deeper potential wells (specifically,
higher escape velocities from the halo center).  Therefore, gravity
more strongly binds the stellar and gaseous contents of such halos
possibly driving more rapid star formation or making the galaxies less
susceptible to feedback mechanisms that suppress star
formation. Higher concentration halos have also formed earlier, on
average \citep[e.g.,][]{wechsler02}. This suggests that any physics
that acts upon the baryonic content of the halo and its high-density
environment may have been operative for a longer time, resulting in a
larger, more luminous galaxy.  Concentration-dependent halo assembly
bias persists to very large separations 
\citep[$r > 10\, h^{-1}\mathrm{Mpc}$, e.g.,][]{wechsler06} and this picture can
therefore accommodate galaxy assembly bias on such large scales.

Another possible explanation for having more luminous galaxies reside
in higher concentration halos is halo-halo interactions.  Recent work
on halo evolution suggests that a more physical definition of the size
of a halo should be $\sim 2-3$ times the traditional virial radius
\citep{wetzel_etal14,adhikari_etal14,wetzel_nagai15,more_etal15,sunayama_etal16}.
This ``splashback radius" incorporates ``splashback" halos that have
passed within the dense environment of the central halo, but are
traditionally classified as distinct halos despite being strongly
influenced by the environment of the central
halo. \citet{wang_etal09}, \citet{wetzel_etal14}, and 
\citet{sunayama_etal16}, all find that these ``splashback" halos are a
significant cause of halo assembly bias on scales $r \sim 1-3\,
h^{-1}\mathrm{Mpc}$, where assembly bias is most pronounced and
strongly scale dependent. Consequently, decorated HOD models exhibit
this same feature \citep{hearin_etal16}, and is not a coincidence that
the greatest improvement of the decorated HOD in the description of
galaxy clustering occurs on these same length scales (see
Fig.~\ref{fig:Mr19samples}-\ref{fig:Mr21samples} and the associated
discussion).

This ``splashback" hypothesis has further merit in that it has the
same sense as suggested by the data. Splashback halos have experienced
mass loss as a consequence of their tidal interactions with a larger,
nearby halo. As a consequence, splashback halos are expected to be
less massive than `regular' host halos that host similar galaxies
\citep{dalal_etal08}.  Meanwhile, high concentrations 
are a hallmark of splashback halos
\citep{sunayama_etal16}. The implication is
a population of halos, correlated on scales $r \sim 1-3\,
h^{-1}\mathrm{Mpc}$, that have concentrations that are high for their
mass and galaxies that are overluminous for their mass. This is
precisely the sense of assembly bias suggested by our analysis.  Note,
though, that this splashback effect cannot explain galaxy assembly
bias on large, linear scales.

Our formulation of the Decorated HOD is not designed to address 
the finer-grained causes for observed correlations. 
Rather, what we have provided in this work is a robust boundary condition 
that more detailed specific models, such as semi-analytical models 
and hydrodynamical simulations, should satisfy. 
However, we reiterate an important corollary to our findings. 
Although SDSS projected clustering measurements favor strong levels of assembly bias, 
the constraints on even our simple $\abias$ parameters are relatively weak. 
This implies that {\em it may not be possible to constrain a finer-grained physical 
picture using projected clustering at a level that is interesting.} 
In order to constrain models that are more complex than simple empirical models 
such as the Decorated HOD, it is necessary to leverage additional constraining power of 
observations beyond two-point clustering.

%------------------------------------------------
\section{Conclusions}
\label{section:conclusions}
%------------------------------------------------

We have re-analyzed the SDSS DR7 measurements of projected galaxy clustering,
$\wprp,$ and number density, $n_{\rm g},$ originally published in \citet{zehavi_etal11}.
The novelty of our work is that we have performed these analyses using the
decorated Halo Occupation Distribution (HOD) formalism, which allows for galaxy
assembly bias and was introduced by \citet{hearin_etal16} precisely for this purpose.
In this work, we provide the first quantitative constraints on
assembly bias derived from the Decorated HOD. Projected galaxy clustering, 
$\wprp(r_{\rm p})$ is already very well-measured within SDSS DR7, therefore it is likely 
that further improvements on assembly bias constraints at low-redshift will require the use of additional 
observables such as galaxy--galaxy lensing, group statistics, void statistics, or numerous other possibilities. 
We are currently exploring many of those possibilities and will report on those results in forthcoming publications. 

Our most important conclusions are as follows.
\ben
\item It is not possible to rule out galaxy assembly bias using SDSS DR7 measurements of $\wprp$ and $n_{\rm g}$.
\item Decorated HOD fits to $\wprp$ and $n_{\rm g}$ favor significant levels of assembly bias, particularly in the lower luminosity thresholds we study. Both the $\magr<-20, -20.5$ samples prefer relatively strong central galaxy assembly bias. The $\magr<-19$ sample favors satellite assembly bias. Assembly bias in these samples improves the capability of the model to describe galaxy clustering near
the one-halo to two-halo transition at $r_{\rm p} \sim 2 \, h^{-1}\mathrm{Mpc}$.
\item The evidence for galaxy assembly bias weakens for brighter galaxy samples: 
Decorated HOD fits to the $\magr<-21$ sample are consistent with both $A_{\rm cen}=0$ and $A_{\rm sat}=0.$ 
This is consistent with the well-established result that
{\em halo assembly bias} weakens with increasing halo mass over the dynamic range relevant to these galaxy samples
\citep[see, e.g., Figure 8 of][and references therein]{hearin_etal16}.
\item Our posteriors and best-fit parameters summarized in
Table 2 supersede the values published in \citet{zehavi_etal11},
as direct-mock-population together with the Decorated HOD allows us to account for
highly significant systematics that have heretofore been neglected from all HOD fits to SDSS data.
We note that our findings update the original \citet{zehavi_etal11} results
{\em even for our fits to standard HODs without assembly bias}, because
the original results were derived from MCMC chains that did not sufficiently sample the HOD model posteriors.
\een

Until such time as we have stringent constraints on the level of assembly bias in the 
universe, accounting for the possibility of assembly bias is not optional in interpreting 
galaxy clustering data. Precise statements about galaxy evolution derived from 
galaxy clustering or attempts to exploit clustering to perform cosmological analyses 
cannot be made robustly without accounting for assembly bias \citep[see][]{zentner_etal14}. 
Together with the work of \citet{hearin_etal16} and the {\tt Halotools} software, we have provided 
an explicit example of one way forward in such analyses. It is our hope that similar analyses 
will be performed on forthcoming galaxy survey data and that these analyses will lead to a
richer understanding of galaxy evolution and the relationship between galaxies and the halos 
and environments in which they are found.

%------------------------------------------------
\section{Acknowledgements}
\label{section:acknowledgements}
%------------------------------------------------

We thank Zheng Zheng for his help in diagnosing the
differences between the present analysis and the SDSS DR7
analysis presented in \citet{zehavi_etal11}. We thank Manodeep 
Sinha for help installing and running the {\tt CorrFunc} package 
as well as comments on an earlier draft of this manuscript. We are grateful to 
Dan Foreman-Mackey and all contributors to the {\tt emcee} and 
{\tt corner} software packages which helped to greatly expedite this 
work. ARZ and APH are grateful to Prince Rogers Nelson for inspiring our creativity. 
The authors gratefully acknowledge the Gauss Centre for Supercomputing
e.V. (www.gauss-centre.eu) and the Partnership for Advanced
Supercomputing in Europe (PRACE, www.prace-ri.eu) for funding the
MultiDark simulation project by providing computing time on the GCS
Supercomputer SuperMUC at Leibniz Supercomputing Centre (LRZ,
www.lrz.de). The Bolshoi simulations have been performed within the
Bolshoi project of the University of California High-Performance
AstroComputing Center (UC-HiPACC) and were run at the NASA Ames
Research Center. The work of ARZ was supported by the US National Science
Foundation via grant AST 1517563. FvdB and JUL are supported by the US National Science
Foundation through grant AST 1516962. APH is funded through the Yale Center 
for Astronomy \& Astrophysics Prize fellowship.

%------------------------------------------------
\bibliography{ms}

%------------------------------------------------
\end{document}